\documentclass[preprint,12pt]{elsarticle}

\usepackage{amssymb}

\usepackage{amsthm}
\usepackage{amsmath}
\usepackage{url} 

\begin{document}

\begin{frontmatter}



\title{CCSNet: a deep learning modeling suite for CO$_2$ storage}

%

\author[label1]{Gege Wen\corref{cor}}
\cortext[cor]{Corresponding author}
\ead{gegewen@stanford.edu}
\author[label1]{Catherine Hay}
\author[label1]{Sally M. Benson}
\affiliation[label1]{organization={Energy Resources Engineering, Stanford University},
            addressline={367 Panama St}, 
            city={Stanford},
            postcode={94305}, 
            state={CA},
            country={USA}}

\begin{abstract}
Numerical simulation is an essential tool for many applications involving subsurface flow and transport, yet often suffers from computational challenges due to the multi-physics nature, highly non-linear governing equations, inherent parameter uncertainties, and the need for high spatial resolutions to capture multi-scale heterogeneity. We developed CCSNet, a general-purpose deep-learning modeling suite that can act as an alternative to conventional numerical simulators for carbon capture and storage (CCS) problems where CO$_2$ is injected into saline aquifers in 2d-radial systems. CCSNet consists of a sequence of deep learning models producing all the outputs that a numerical simulator typically provides, including saturation distributions, pressure buildup, dry-out, fluid densities, mass balance, solubility trapping, and sweep efficiency. The results are 10$^3$ to 10$^4$ times faster than conventional numerical simulators. As an application of CCSNet illustrating the value of its high computational efficiency, we developed rigorous estimation techniques for the sweep efficiency and solubility trapping. 
\end{abstract}

\end{frontmatter}


\section{Introduction}
\label{Introduction}
Multiphase flow in porous media is important for many subsurface flow and transport problems such as hydrocarbon production~\cite{aziz1979petroleum} and carbon capture and storage (CCS)~\cite{pachauri2014climate}. Numerical simulation is the primary tool used for predicting field-scale multiphase flow by solving spatially and temporally discretized mass and energy balance equations~\cite{allen1985numerical, chierici1995simulation, pruess2005eco2n}. However, numerical simulation for multiphase flow problems is computationally expensive due to the multiphysics problem nature~\cite{khebzegga2020continuous}, highly nonlinear governing partial differential equations (PDEs)~\cite{orr2007theory}, multiscale heterogeneity in the permeability field~\cite{pini2012capillary}, and need for high spatial resolution of the grids~\cite{Doughty2010, Wen2019}. The inherent uncertainty in the subsurface geology necessitates probabilistic assessments and history matching~\cite{Kitanidis2015}, which often require prohibitively large numbers of simulation runs. To aid engineering decisions, `surrogate' models with lower fidelity but greater computational efficiency are often developed for specific tasks~\cite{cardoso2009development, razavi2012review, bazargan2015surrogate, hamdi2017gaussian, tian2017gaussian}.

Here we propose a deep learning approach for solving subsurface flow and transport problems with the fidelity of a traditional simulator and the speed of surrogate models or even faster. Unlike previous surrogate methods that are often developed on a `task' basis~\cite{cardoso2009development, razavi2012review, bazargan2015surrogate, hamdi2017gaussian, tian2017gaussian, tang2020deep, jin2020deep, mo2019deep, zhong2021deep}, we demonstrate a deep learning tool, CCSNet, which can provide solutions to an entire class of multiphase flow problems, namely, CO$_2$ storage problems. CCS is a climate change mitigation technology that requires injection of supercritical CO$_2$ into saline aquifers for long term storage~\cite{IEA2020ccus}. CCSNet can solve for nearly all realistic scenarios that entail injecting CO$_2$ into a 2d-radial system through a vertical injection well~\cite{yamamoto2011investigation}.  In such systems, the complex interplay between capillary, gravity, and viscous forces controls the migration of CO$_2$~\cite{Yamamoto2011, Saadatpoor2010, Krevor2015, Wen2019}. CO$_2$ migrates horizontally away from the injection well due to viscous forces while rising upwards due to to gravitational forces. Subsurface geological heterogeneity results in variations of permeability and capillary entry pressure~\cite{Ide2007, pini2012capillary}, which have a first-order effect on plume migration patterns, pressure buildup, trapping, and sweep efficiency~\cite{Wen2019}. Accurately modeling of these phenomena requires numerical simulations with high spatial and temporal resolutions~\cite{Pruess2011, Doughty2010}, making rigorous probabilistic assessments, optimization, and history matching for CO$_2$ storage especially computationally intensive using conventional numerical simulators. 

Deep learning has recently shown a growing potential for applications to subsurface flow and transport problems~\cite{Zhu2019, jin2020deep, mo2019deep, tang2020deep, fuks2020physics, zhong2021deep, WEN2021103223, wu2020physics, he2020physics,liu2020petrophysical, jiang2021deep}. Physics informed or physics constrained machine learning approaches encode governing PDEs in the loss function and solve the problem through automatic differentiation~\cite{ Zhu2019,wu2020physics, he2020physics, haghighat2021sciann}. To date, physics-informed machine learning models have not been successful in providing accurate approximations for hyperbolic PDEs that govern most multiphase flow problems~\cite{fuks2020physics}. Supervised learning approaches use data generated by numerical simulators to train deep learning models: these have shown encouraging results for specific uncertainty quantification or history matching tasks~\cite{mo2019deep, tang2020deep, Zhong2019}. In fact, supervised learning models can represent any complicated relationship given sufficient data and adequate training~\cite{haykin2010neural} and we develop CCSNet based on this principle. We demonstrate in this paper that deep learning tools have functionalities beyond merely used as task driven surrogate models. Instead, CCSNet provides solutions to a whole class of problems -- in essence, for certain applications providing an alternative to conventional numerical simulation.

\begin{figure}[t!]
\centering
\includegraphics[width=\textwidth]{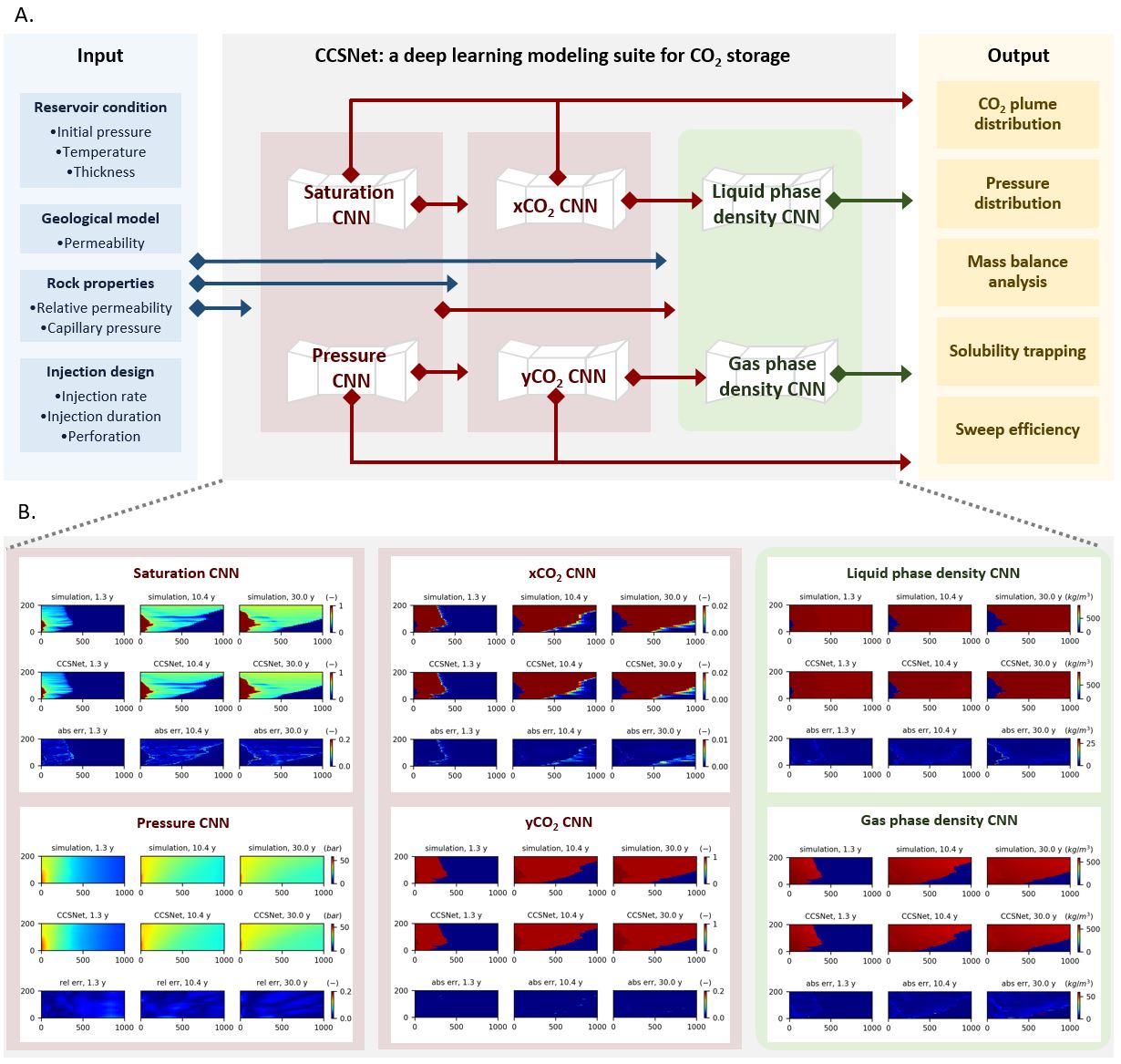}
\caption{A. CCSNet's inputs, prediction sequence, and outputs. The input section illustrates the four variable categories and specific variables in each categories. The prediction sequence section shows the 6 convolutional neural network (CNN) models. The output section shows the variables that CCSNet can produce. The arrows indicate the specific input and output for each model in the prediction sequence. B. Comparisons of the numerical simulation outputs, outputs predicted by CCSNet, and absolute/relative error at three arbitrary time snapshots for each model. The figures lie on the $(r,z)$ coordinate. The $r$ direction can extend to 100,000 m and the examples shown here are cropped.}
\label{fig:workflow}
\end{figure}

A major challenge for developing general-purpose tools for classes of problems is to design and create a training set that can fully represent the problem domain. Here we train CCSNet with a data set containing highly resolved and full-physics numerical simulation outputs that are representative of all realistic scenarios for 2d-radial CO$_2$ injection, including extensive ranges of reservoir conditions, fluid properties, geological attributes, rock properties, multiphase flow properties, and injection designs. Figure~\ref{fig:workflow}A shows the sequence of convolutional neural network (CNN) models in CCSNet that collaboratively provide predictions of salient outputs from conventional numerical simulators, namely, CO$_{2}$ gas saturation distribution, pressure buildups, the molar fractions of CO$_2$ and fluid densities for gas and liquid phases~\cite{Pruess1999}. The full set of outputs allows us to evaluate how well the results satisfy the governing conservation equations without explicitly representing them in the loss function. CCSNet is nearly as accurate as numerical simulation for all realistic cases in the problem domain while being 10$^3$ to 10$^4$ times more computationally efficient. To demonstrate the value of CCSNet's high computational efficiency, we used stochastic sampling of the problem domains to develop an estimation technique for sweep efficiency and solubility trapping, two of the important considerations when selecting sites for CCS projects.

\section{Methodology}
\label{Methodology}
This section describes the governing equations, training data set generation, model architecture, data configuration, and training strategy of CCSNet.

\subsection{Governing equations}
\label{GoverningEquations}
For the CO$_2$ and water multiple-phase flow problem, 
the general form of mass accumulation for component $\kappa= CO_2$ or $water$ is written as~\cite{Pruess1999}:
\begin{equation}
\frac{\partial M^\kappa}{\partial t}=-\nabla\cdot \mathbf{F}^\kappa+q^\kappa,
\end{equation}
For each component $\kappa$, the mass accumulation term $M_{\kappa}$ is summed over phases $p$,
\begin{equation}
M^\kappa=\phi\sum_pS_p\rho_pX^\kappa_p,
\label{mass_balance}
\end{equation}
where $\phi$ is the porosity, $S_p$ is the saturation of phase $p$, $\rho_p$ is the density of phase $p$, and $X^\kappa_p$ is the mass fraction of component $\kappa$ presents in phase $p$. For each component $\kappa$, we also have the advective mass flux $\mathbf{F}^\kappa|_{adv}$ obtained by summing over phases $p$,
\begin{equation}
\mathbf{F}^\kappa|_{adv}=\sum_{p}X^\kappa_p\mathbf{F}_{p}
\end{equation}
where each individual phase flux $\mathbf{F}_{p}$ is governed by Darcy's law:
\begin{equation}
\mathbf{F}_p = \rho_p \mathbf{u}_p = -k\frac{k_{r,p}\rho_p}{\mu_p}(\nabla P_{p} - \rho_p\mathbf{g}).
\end{equation}
Here $\mathbf{u}_p$ is the Darcy velocity of phase $p$, $k$ is the absolute permeability, $k_{r,p}$ is the relative permeability of phase $p$, $\mu_p$ is the viscosity of phase $p$, and $\mathbf{g}$ is the gravitational acceleration. The fluid pressure of phase $p$
\begin{equation}
P_p = P + P_{c}
\end{equation}
is the sum of the reference phase (usually the gas phase) pressure $P$ and the capillary pressure $P_{c}$. To simplify the problem setting, our simulation does not explicitly include molecular diffusion and hydrodynamic dispersion.

\subsection{Training data set generation} 
We used the numerical simulator ECLIPSE (e300) to generate a large data set that is representative of most potential scenarios for  CO$_2$ storage in deep geological formations. ECLIPSE is a state-of-the-art full physics numerical simulator that uses the finite difference system with upstream weighting and the adaptive IMplicit method for simulation~\cite{eclipse}. The modeled volume is a radially symmetrical cylindrical volume that is 200m thick and 100,000m along the radius. The reservoir has no-flow boundaries on the top and bottom; the large radius mimics an infinite acting boundary on the radial direction. This geometry represents CO$_2$ injection into a regional-scale saline formation with a negligible dip, such as found in the Illinois Basin and parts of the North Sea and Gulf Coast. The modeled volume is isothermal and contains pure water prior to CO$_2$ injection. The vertical injection well is located at the center of the modeled volume, and the well radius is 0.1m. The injection well has no cross-flow, which means CO$_2$ can flow only from the well to the reservoir. The well has a single and continuous perforation and injects at a constant rate.

We used 96 uniform grid cells in the vertical direction and 200 gradually coarsened grid cells in the radial direction to represent the reservoir. This grid design is sufficiently refined to resolve the plumes in heterogeneous reservoirs while it remains computationally tractable for the purpose of training the deep learning models~\cite{Wen2019}. The numerical simulation runs for 30 years with 24 gradually coarsening time snapshots. Details of the temporal and spatial grid are discussed in \ref{GridResolution}. 

For each simulation case, we sample the inputs from the following four main categories.

\subsubsection{Reservoir conditions} This category consists of formation thickness, initial pressure, and temperature, which are the most basic types of information available for any geological formation. 

Existing machine learning-based methods for predicting subsurface flow problems usually suffer from fixed data dimensions, which significantly limits the models' applicability. To account for the variable formation thicknesses, we assign extremely low permeability (10$^{-7}$mD) to layers in excess of the actual reservoir thickness. Using this method, CCSNet can handle formation thicknesses from 15m to 200m, which covers most of the known CO$_2$ storage projects operating today~\cite{gccsidatabase}. In future revisions, thicker reservoirs can also be included. The initial pressure and temperature in a formation depend on the depth and geothermal gradient. Formations that are too shallow are unsuitable for injection because CO$_2$ might not be in a super-critical state under reservoir condition; for formations that are too deep, drilling costs are prohibitively high for CO$_2$ storage~\cite{NationalAcademiesofSciencesEngineering2018}. Therefore, to generate realistic combinations of initial pressure and temperature, we first randomly sample the reservoir initial pressure from 100 to 300 bar, which corresponds approximately to formation depth from 1,000 to 3,000m. Subsequently, for the reservoir temperature, we sampled the geothermal gradient from 18 to 50 C$^\circ$/km and created a wide range of temperature values from 35 to 170$^\circ$C. 

\subsubsection{Geological model} 
The geological model describes the spatial distribution of permeability values. We train CCSNet with a data set containing various types of permeability maps. The permeability maps representing different depositional environments include a broad range of permeability values (10$^{-3}$ mD to 10$^2$ D), a wide variety of horizontal and vertical correlations, and various permeability distributions, such as Gaussian, non-Gaussian, bi-modal, multi-modal distribution, and uniform distributions  (statistical characteristics summarized in \ref{Statistical}). 

\subsubsection{Rock properties} 
Commonly used rock property characteristic curves for CO$_2$ storage include relative permeability curves and capillary pressure curves. Both characteristic curves substantially impact the rate and direction of plume migration and are incorporated into CCSNet. To account for different relative permeability and capillary pressure curves, we sample the irreducible water saturation and van Genuchten function scaling factor according to references of rock types that can be used for CO$_2$ storage~\cite{krevor2012relative, benson2013relative}. The irreducible water saturation controls the relative permeability and capillary pressure characteristic curves. We used Corey's curves to model relative permeability curves as a function of water phase saturation ($S_{w}$):
\begin{equation}
\begin{split}
k_{r,w}&={S_w^*}^{n_w},\\
k_{r,CO_2} &=  k_{r,CO_2}(S_{wi})(1-{S_w^*})^2[1-({S_w^*})^{n_{CO_2}}],
\end{split}
\end{equation}
where $k_{r,w}$ is the relative permeability of the water phase, $k_{r,CO_2}$ is the relative permeability of the CO$_2$ phase,  $S_{wi}$ is irreducible water saturation, coefficient $n_w=6$, coefficient $n_{CO_2}=5$, coefficient $k_{r,CO_2}(S_{wi})=0.95$, and $S_{l}^*=(S_l-S_{wi})/(1-S_{wi})$. To create different sets of relative permeability curves, we sampled $S_{wi}$ from 0.1 to 0.3 in the training set.


The capillary pressure curves are modeled by the van Geneuchten function:
\begin{equation}
P_{c}=P_{e}[(S^*)^{-1/\lambda}-1]^{1-\lambda},
\label{eqs:lambda}
\end{equation}
where $P_c$ represents capillary pressure, $P_e$ represents capillary entry pressure, and $S^*=(S_w-S_{wi})/(S_{ls}-S_{lr})$. Note that here we used an approximation of $S_{lr}=0.999$ to represent the capillary entry pressure to avoid numerical errors in ECLIPSE. In the data set, we randomly sampled the scaling factor $\lambda$ from 0.3 to 0.7 to create capillary pressure curves with different slopes. The capillary entry pressure is scaled according to the permeability in each grid cell by Leverett J-function:
\begin{equation}
P_{e}=\frac{\sqrt{k_{ref}/\phi_{ref}}}{\sqrt{k/\phi}}P_{ref},
\end{equation}
where $k_{ref} = 3.95 \times 10^{-15}$ $m^2$, $\phi_{ref}=0.185$, and $P_{ref}=7,500 Pa$.
For the van Genuchten curve, the $S_{wi}$ is the same as in Corey's curve.

\subsubsection{Injection design} 
We created various combinations of injection rates, injection depths, and perforation thicknesses. In the training set, the maximum injection rate is 2 MT/year and the minimum injection rate is 0.02 MT/year. In addition to the injection rates, injection locations and perforation thicknesses also significantly influence the plume migration, especially in heterogeneous reservoirs. We created a wide range of injection strategies with the injection perforation interval thicknesses range from 15m to 200m; the top of the perforated interval is placed randomly within the depth interval of the injection well.

\subsection{Model architectures}
\begin{figure}[!ht]
\includegraphics[width=\textwidth]{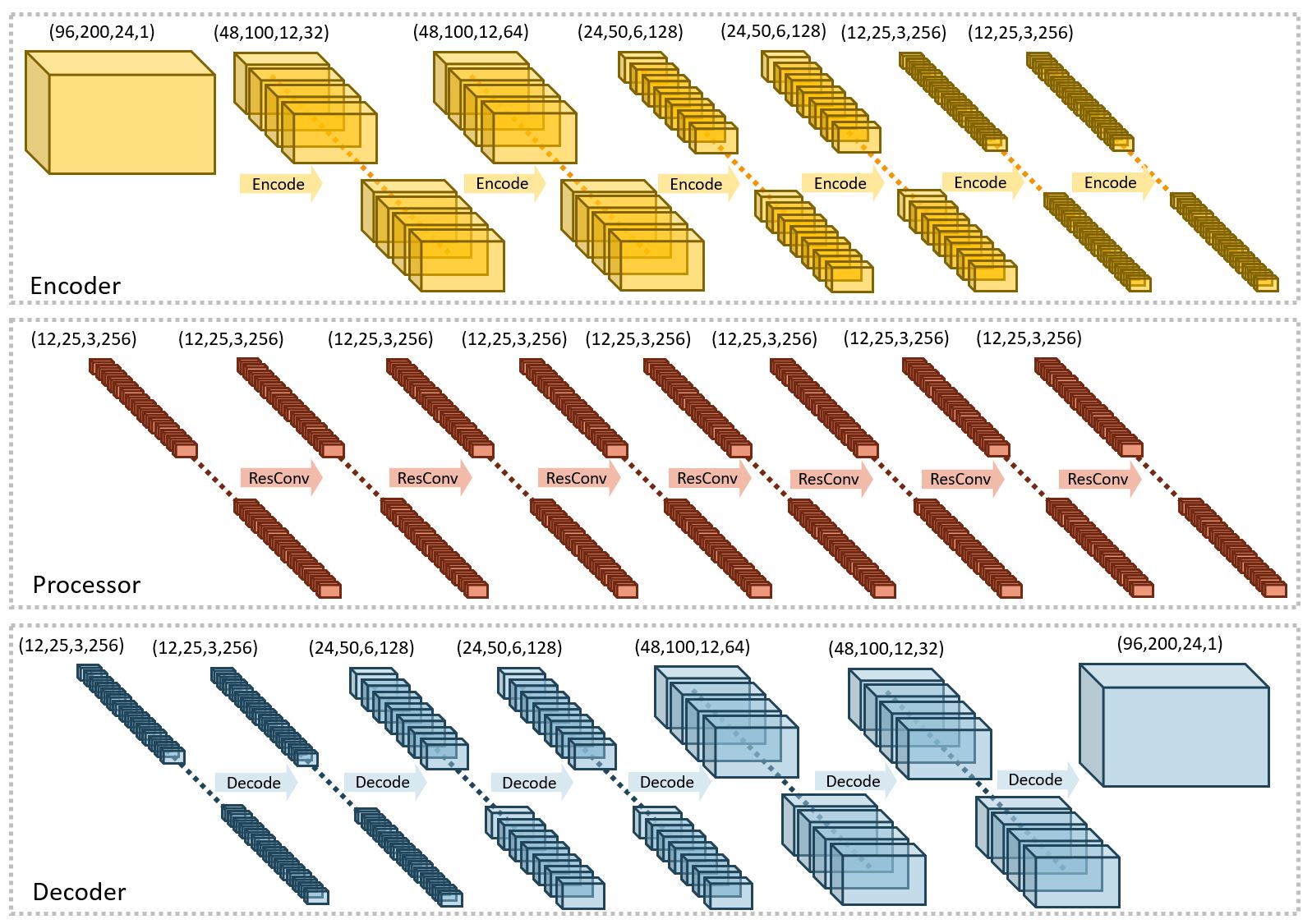}
\caption{Model schematics of the Saturation CNN. The \textbf{Encode} operation consists of \texttt{Conv3D}/\texttt{BN}/\texttt{ReLu}; the \textbf{ResConv} operation consists of \texttt{Conv3D}/\texttt{BN}/\texttt{Conv3D}/\texttt{BN}/\texttt{ReLu}/\texttt{Add}; the \textbf{Decode} operation consists of \texttt{UnSampling}/\texttt{Padding/Conv3D/BN/Relu} . The last dimension in the bracket denotes the number of channels.}
\label{fig:model}
\end{figure}

We designed a temporal-3d CNN model architecture for predicting the dynamic changes of CO$_2$ gas saturation, pressure buildup, molar fractions, and densities in each phases. The temporal-3d CNN consists of 3d convolutional kernels~\cite{tran2015learning} that can extract information in both the temporal and spatial dimensions, which are adopted from state-of-the-art video classification and human action recognition models~\cite{song2017end, xie2018rethinking}. Notably,  we trained the temporal-3d CNN on data that has both non-uniform spatial and temporal dimension. Our results show that the temporal-3d CNN has excellent performance in non-uniform spatial-temporal systems, which significantly improved the models' applicability. For the input/output regression mapping, we used an encoder-decoder structure that contains three major components: encoder, processor, and decoder (Figure~\ref{fig:model}). 

The encoder maps the input 3d-volume to the input feature embedding. The processor learns the relationship between the input's embedding and the output's embedding using multiple 3d-ResConv blocks that we designed based on the well-known 2d-residual learning block~\cite{he2016deep}. The decoder projects the embedding of the output to the temporal-3d output space that represents the dynamic change of saturation, pressure, and dissolved phase molar fraction. Our work shows that the temporal-3d encoder-decoder architecture has performance  superior to that of the U-Net based architectures~\cite{ronneberger2015u} because, we hypothesize, the input and output exist in different spatial and temporal spaces. 

The schematic in Figure~\ref{fig:model} shows the network depth, and sizes of the Saturation CNN. For pressure, molar fractions, and densities in each phases, the depths and the sizes of the temporal-3d model architecture were  optimized to provide accurate predictions each specific output. Parameters for each model are summarized in \ref{app:SaturationCNN}, \ref{app:PressureCNN}, and \ref{app:xco2CNN}.

\subsection{Data configuration and augmentation} 
\subsubsection{Outputs}
Numerical simulation outputs at any arbitrary time step can be represented as 2d matrices in the dimension of $96\times200$ $(r,z)$. The 2d matrices are stacked along the temporal dimension to construct the temporal-3d volume with the dimension of $96\times200\times24$ $(r,z,t)$. The output data of gas saturation distribution, pressure buildup, molar fraction of CO$_2$ in the liquid phase (xCO$_2$) and gas phase (yCO$_2$), and densities in each phase are all configured in this manner. 

To improve the training efficiency, we applied min-max normalization to the output values of pressure buildup and xCO$_2$. We also applied a data augmentation technique in addition the min-max normalization for the outputs of yCO$_2$ and densities in each phase. For these outputs, the magnitude of the output values within the plume have a large difference comparing to the values outside the plume. For example, gas densities within a plume has magnitude of few hundreds ($kg/m^3$) with a small variation; gas density outside of a plume is always zero. Therefore, simply applying a min-max normalization to these outputs would suppress the details within the plume area. We used the data augmentation technique that casts the values outside the plume to be a constant slightly smaller than the minimum value within the plume. This technique allows us to maintained the details within the plume and produces highly accurate prediction for yCO$_2$ and densities in each phase.


\subsubsection{Inputs}
The inputs to the CNN models are designed to have the identical shape as the outputs. The high dimensional volume ($96\times200\times24$) provides room for incorporating all of the aforementioned input variables: reservoir conditions, geological attributes, rock properties and injection design. For each input, the permeability map and reservoir thickness are represented in a $96\times200$ matrix. The injection perforation location is represented by a binary matrix where only grid cells next to the perforation interval are marked by one. The variable of initial pressure, temperature, injection rate, irreducible water saturation, and van Genuchten scaling factor are scalar values which we broadcast into matrices in the dimension of $96\times200$. These matrices are concatenated to construct the input volume in dimension of $96\times200\times24$. Notice that these input variables only populate 7 of the 24 slices available in the input volume. The idle slices are populated with permeability maps here and can be converted easily to directional permeability or porosity in the future. 

The Saturation and Pressure CNNs both use this input volume as their training input. For the xCO$_2$ and yCO$_2$ CNNs, the prediction also requires the predicted gas saturation and pressure buildup in addition to the input volume. Therefore, we concatenated the gas saturation volumes, pressure buildup volumes, and the original input volumes to create a 4d training input with the dimension of $96\times200\times24\times3$. Similarly, for training the CNNs that predict the densities in each phase, we constructed a similar 4d input that consists of the formation temperature, pressure, and the molar fraction in the specific phase.

The training/validation data split is 10/1 with 19,000 training samples and 1,900 validation samples. 

\subsection{Training strategy}
The loss function that we used is the Mean Square Error (MSE) loss: $$L_{MSE}=\frac{1}{N}\sum^N_{i=1}||\mathbf{y}_i-\hat{\mathbf{y}}_i||^2_2,$$ where $N$ is the number of training samples in a batch, $\mathbf{y}$ is the true output in the data set,  $\hat{\mathbf{y}}$ is the output predicted by the temporal-3d CNN model, described as $\hat{\mathbf{y}}_i=\mathbf{f}(\mathbf{x}_i,\boldmath{\theta})$, where $\boldmath{\theta}$ is the model's learnable parameters and $\mathbf{x}_i$ is the input. During training, $\boldmath{\theta}$ is updated based on the gradient to the loss function with respect to the $\boldmath{\theta}$ (also referred to as back propagation in machine learning). We used the Adam optimizer~\cite{kingma2014adam} for the minimization of the loss function. The Glorot normal initializer~\cite{glorot2010understanding} (also referred to as the Xavier normal initializer) was used to initialize the convolutional layers' kernels in the CNNs. We applied L2 weight regularizers on the convolutional layers with a hyperparameter of 0.001 to reduce overfitting. The learning rate was initialized to be $10^{-4}$ and manually reduced to $10^{-7}$ throughout the training process. Our previous experiments show that the training efficiency of the CNNs is nearly insensitive to the choice of batch size~\cite{WEN2021103223}.  The models were trained on NVIDIA v100 GPUs, and the training duration varied from a few days to a week. 

\section{Results}
\label{result}

\begin{table}[!t]
\centering
\footnotesize
\caption{Accuracy summary for model/output. All values were evaluated base on 1,000 randomly chosen samples. $R^2$ refers to the average scores in the training or validation set, $\mu$ refers to a mean, $\sigma$ refers to a standard deviation, $PAE$ refers to absolute errors within the plume; $RAE$ refers to relative absolute errors for the whole field; $PRE$ refers to relative errors within the plume.}
\begin{tabular}{p{0.23\linewidth}rrrr}
Model / Output & Metric & Training & Validation & Unit \\
\hline
Saturation CNN & $R^2$ &  0.999 & 0.998  & -\\
               & $\mu_{PAE}$ & 0.008 & 0.009  & $m^3/m^3$\\
               & $\sigma_{PAE}$ & 0.013 & 0.014  & $m^3/m^3$\\
\hline
Pressure CNN   & $R^2$ & 0.997 & 0.996 & - \\
               & $\mu_{RAE}$ & 2.3 & 2.5 & \%  \\
               & $\sigma_{RAE}$ & 1.1 & 1.2  & \%\\
\hline
xCO$_2$ CNN    & $R^2$ & 0.998  & 0.998 & - \\ 
               & $\mu_{PAE}$ & 1.58$\times10^{-4}$ & 1.69$\times10^{-4}$ & $mol/mol$ \\ 
               & $\sigma_{PAE}$ & 6.60$\times10^{-5}$ & 8.31$\times10^{-5}$ & $mol/mol$ \\ 
\hline
yCO$_2$ CNN    & $R^2$ & 1.000  & 1.000   & -\\ 
               & $\mu_{PAE}$ &  6.80$\times10^{-4}$ &  8.09$\times10^{-4}$  & $mol/mol$\\ 
               & $\sigma_{PAE}$ & 3.42$\times10^{-4}$  &  4.41$\times10^{-4}$  & $mol/mol$\\ 
\hline
Liquid phase   & $R^2$ &  1.000   & 1.000   & - \\ 
density CNN    & $\mu_{PRE}$ &  0.05  &  0.06 & \% \\ 
               & $\sigma_{PRE}$ &  0.06  &  0.06  & \% \\ 
\hline
Gas phase      & $R^2$ &  1.000 &  1.000 & - \\
density CNN    & $\mu_{PRE}$ &  0.01  & -0.01  & \% \\
               & $\sigma_{PRE}$ & 0.14   &  0.16 & \% \\
\hline
Mass balance 
 & $R^2_{liq}$ & 0.999 & 0.999 & - \\
 & $\mu_{liq}$ &  0.06 & 0.07 & \% \\
 & $\sigma_{liq}$ &  0.68 & 1.07 & \%  \\
 & $R^2_{gas}$ & 1.000 & 1.000 & - \\
 & $\mu_{gas}$ &  0.07 & 0.07 & \% \\
 & $\sigma_{gas}$ &  0.76 & 0.93 & \%  \\
 & $R^2_{total}$ & 1.000 & 1.000 & - \\
 & $\mu_{total}$ & -0.09 & 0.08 & \% \\
 & $\sigma_{total}$ & 0.74  & 0.85 & \%  \\
 \hline
\end{tabular}
\label{table}
\end{table}

\subsection{CO$_2$ gas saturation distribution}

\begin{figure}[!t]
\centering
\includegraphics[width=\textwidth]{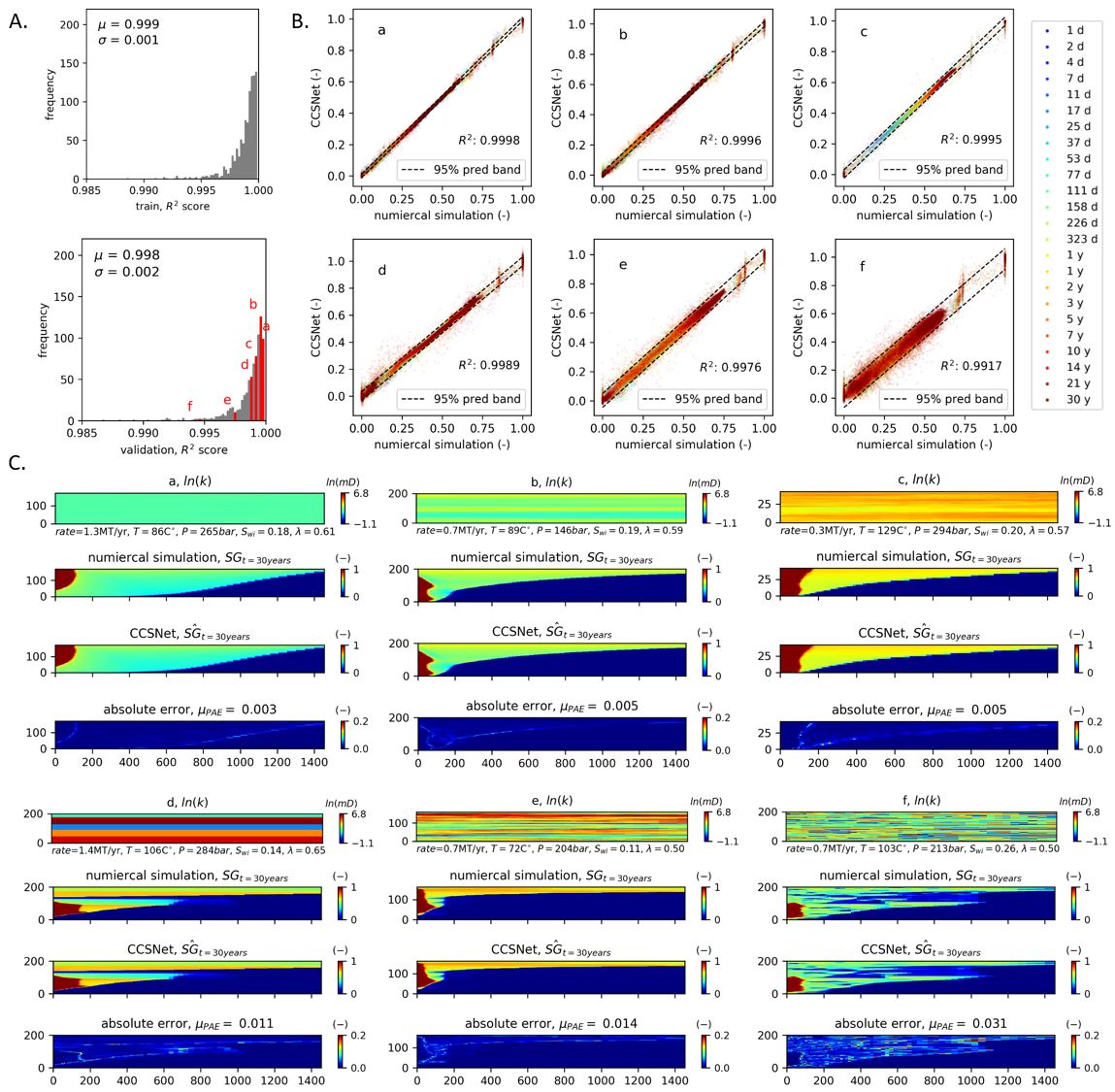}
\caption{A. Histograms of Saturation CNN's R$^2$ scores in the training and the validation set with the mean and standard deviation. On the validation set histogram, the red bars denote the R$^2$ score of the 6 examples in B and C. The alphabetical order corresponds to cases with R$^2$ scores at the 99, 95, 70, 30, 5, and 1 percentile. B. Gas saturation predicted by CCSNet vs. numerical simulation on each grid for the 6 examples. The colors of the points represent the injection duration. C. The permeability map, numerical simulation output, CCSNet predicted output, and absolute error for the 6 examples at 30 years. Inputs variables are summarized under each permeability map.  The horizontal axes indicate the radial direction and the plume can extend out of the plot area. The mean absolute error within the plume ($\mu_{PAE}$) is shown for each example. The reservoir thicknesses are marked on the vertical axes. }
\label{fig:scatter_sg}
\end{figure}

Using the temporal-3d CNN model illustrated in Figure~\ref{fig:model}, we trained the Saturation CNN to predict dynamic CO$_2$ gas saturation distributions as a function of space and time. Given information about the reservoir conditions, geological attributes, rock properties and injection patterns, the Saturation CNN generates predictions of dynamic CO$_2$ gas saturation distributions in $\sim$0.05s, which is more than 10$^4$ times faster than conventional numerical simulators (details on computing specifications in discussed in Section~\ref{efficiency}).

Figure~\ref{fig:workflow}B shows an example of the Saturation CNN's prediction at several time snapshots in comparison with the numerically simulated output. This example demonstrates the multi-physics nature of the problem: the effects of viscous forces due to injection; the effects of gravity that lead to buoyancy induced flows; and spatially varying rock properties that locally counteract buoyancy. A dry-out zone also forms near the injection perforation, where the liquid-phase water vaporizes entirely into the gas phase. Although the dry-out is challenging for most numerical simulators due to the sharp gas saturation gradient, the Saturation CNN accurately predicts this in addition to the saturation variations caused by geological heterogeneity. 

Based on a 1,000 randomly chosen examples, we show that the Saturation CNN is highly accurate (Table~\ref{table}, Figure~\ref{fig:scatter_sg}A). Similarly high $R^2$ values in the training and validation set demonstrate that the model has successfully learned the underlying relationship between the input parameters and the corresponding saturation plume behavior instead of merely memorizing the training data. We use six examples in Figure~\ref{fig:scatter_sg}B to demonstrate the prediction performance at different $R^2$ scores (ranked from high to low), in which we show that the Saturation CNN performs equally well throughout the prediction period and over the entire range of saturation values. Cases with smoother permeability maps are matched almost perfectly. Even for the most challenging example that is in the lowest 5\% of the validation set, the predicted saturation distribution is in close agreement to the simulated output. The accuracy statistics provided in Table~\ref{table} indicate that the Saturation CNN provides predictions that are sufficiently accurate for predicting plume migration, sweep efficiency assessment, plume footprint prediction, and risk analysis. 

\subsection{Pressure buildup}
\begin{figure}[!t]
\centering
\includegraphics[width=\textwidth]{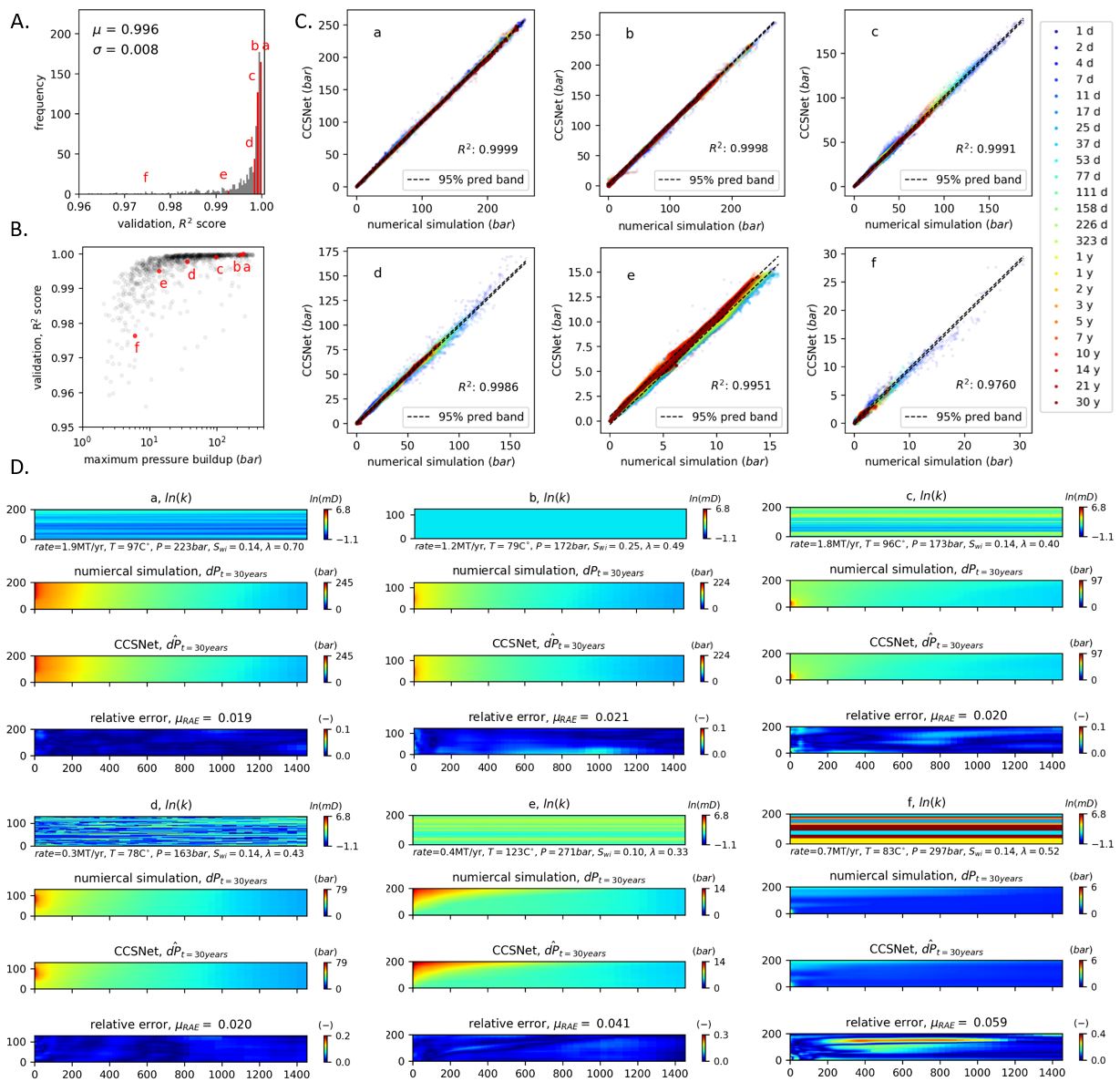}
\caption{A. Histogram of Pressure CNN's R$^2$ scores in the validation set with the mean and standard deviation. The red bars mark the score of the 6 examples in B, C and D. The alphabetical order corresponds to cases with R$^2$ scores at the 99, 95, 70, 30, 5, and 1 percentile. B. Scatter plot of the $R^2$ scores vs. average pressure buildup. C. Pressure buildup predicted by CCSNet vs. numeral simulation on each grid for the 6 examples. The colors of the points represent the injection duration. D. The permeability map, numerical simulation output, CCSNet output, and relative error for the 6 examples at 30 years. Inputs for injection rate, temperature, initial pressure, irreducible water saturation, and capillary pressure scaling factor are summarized under the permeability map. The mean absolute relative error ($\mu_{RAE}$) is shown for each case. The horizontal axes indicate the radial direction and the reservoir thicknesses, are marked on the vertical axes.}
\label{fig:scatter_pressure}
\end{figure}
We trained another temporal-3d CNN model, Pressure CNN (model parameters summarized in \ref{app:PressureCNN}), to predict the pressure buildup due to CO$_2$ injection. The Pressure CNN can be used independently from the Saturation CNN. Pressure buildup predictions take $\sim$0.04s. Figure~\ref{fig:workflow}B shows an example of the Pressure CNN's output at various times, in which the CO$_2$ is injected in the lower half of the reservoir, creating a zone of high pressure buildup near the injection perforation. 

Pressure buildups vary widely in the the reservoir and between cases, ranging from almost 400 bars near the injection well in over pressured reservoirs, to 0 bars near the reservoir boundary in highly permeable reservoirs. The Pressure CNN successfully accounts for the distinctly different pressure behaviors in different reservoirs and produces highly accurate predictions (Table~\ref{table}, Figure~\ref{fig:scatter_pressure}A). The six examples in Figure~\ref{fig:scatter_pressure}C and D illustrate the excellent performance in predicting dynamic propagation of the pressure buildup in the reservoir throughout the injection period. Unlike saturation predictions where permeability heterogeneity dominants the performance, the Pressure CNN's performance is correlated to the magnitude of pressure buildups, and poorest performance is for cases with the smallest pressure buildups (Figure~\ref{fig:scatter_pressure}B). 

\subsection{Mass balance analysis}

Ability to accurately track the total mass balance and distribution of mass between phases is a critical measure of model performance and allows us to evaluate how well the deep learning outputs satisfy the governing conservation laws without explicitly representing the PDEs in the loss function. CCSNet uses six deep learning models to predict all the components needed to perform a mass balance (Figure~\ref{fig:workflow}A). The following equation describes the CO$_2$ mass balance in the reservoir at a given time step:
\begin{equation}
M=\sum_{i,n}V_{n}(\phi S\rho X)_{i,n}
\end{equation}
where $M$ is the total CO$_2$ mass, $i$ denotes the phase (gas or liquid), $n$ denotes the spatial grid,  $V$ is cell volume, $\phi$ is porosity, $S$ is saturation, $\rho$ is density, and $X$ is the mass fraction of CO$_2$. 

CCSNet generates each variable required in the mass balance analysis for the injected CO$_2$. The Saturation CNN provides $S_{i,n}$. Since the rock is compressible, the Pressure CNN is used together with the compressibility of the rock to predict $\phi_{i,n}$. Additionally, we developed and trained a model for predicting molar fractions of CO$_2$ in the liquid (xCO$_2$ CNN), a model for predicting molar fractions of CO$_2$ in the gas phase (yCO$_2$ CNN), as well as two models for predicting densities of the liquid and gas phases. Examples of each model's output are shown in Figure~\ref{fig:workflow}B. Refer to \ref{MassBalance} for details on the mass balance calculations. 

\subsubsection{xCO$_2$ CNN} 
Prediction of the molar fraction of dissolved CO$_2$ in the liquid phase (xCO$_2$) requires information about the gas saturation distribution, temperature, and pressure. CO$_2$ dissolves into the reservoir fluid wherever separate phase CO$_2$ is present. For the two-component system studied here, pressure and temperature control the solubility of CO$_2$. A small amount of dissolved CO$_2$ migrates in advance of the plume. No dissolved CO$_2$ appears in the dry-out zone near the injection well because the entire liquid phase is vaporized into the gas phase and transported away from the dry-out zone. Taking these factors into account, we use predicted outputs from the Pressure CNN and Saturation CNN in addition to the original input to train the xCO$_2$ CNN. Our experiments show that using this concatenated input significantly reduced over-fitting comparing to training with the original input. 

\begin{figure}[ht]
\centering
\includegraphics[width=\linewidth]{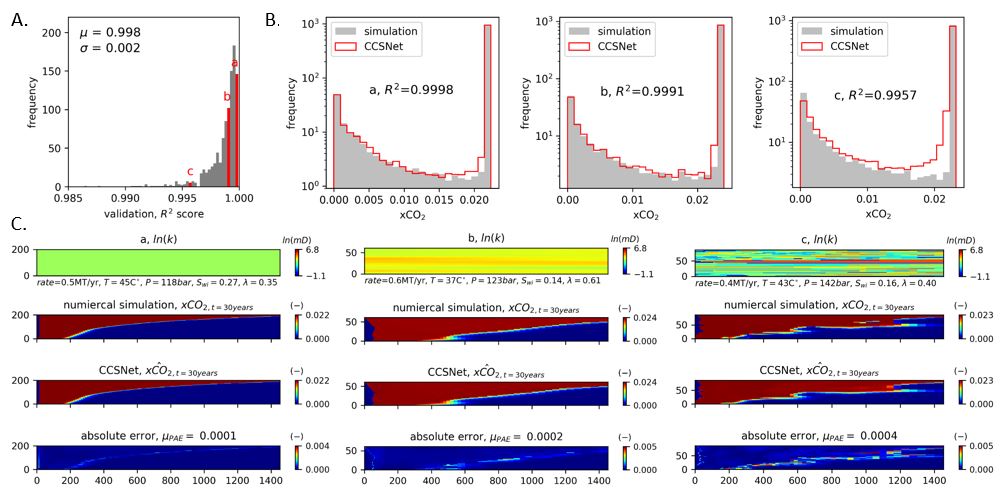}
\caption{A. Histograms of xCO$_2$ CNN's R$^2$ scores in the training and the validation set with mean and standard deviation. The red bars mark the score of the 6 examples in B and C. The alphabetical order corresponds to cases with R$^2$ scores at the 95, 50, and 5 percentile. B. Histogram comparisons between the numerical simulation's output and CCSNet's output within the plume for 3 examples. C. The permeability map, numerical simulation output, CCSNet output, and absolute error for 3 examples at 30 year. Inputs for injection rate, temperature, initial pressure, irreducible water saturation, and capillary pressure scaling factor are summarized under the permeability map. The mean absolute error within the plume ($\mu_{PAE}$) is shown for each example. The horizontal axes indicate the radial direction, and the reservoir thicknesses are marked on the vertical axis.}
\label{fig:xco2}
\end{figure}

The trained xCO$_2$ CNN performs very well (Table~\ref{table}). Three examples in Figure~\ref{fig:xco2}B and C demonstrate the performance of the xCO$_2$ CNN. The prediction of xCO$_2$ CNN is more accurate in relatively homogeneous reservoirs where the dissolved phase xCO$_2$ stays close to the saturation plume front (Figure~\ref{fig:xco2}C.a). Heterogeneous cases such as Figure~\ref{fig:xco2}C.c are more challenging because dissolved CO$_2$ migrates in advance of the plume at various velocities. 

\subsubsection{yCO$_2$ CNN} 
The molar fraction of CO$_2$ in the gas phase depends on temperature and pressure. A small fraction of water vaporizes into the gas phase except in the dry-out zone, where the gas-phase contains nearly purely CO$_2$. We trained the yCO$_2$ CNN to predict the molar fraction of CO$_2$ in the gas phase given the temperature, gas saturation predicted by Saturation CNN, and pressure predicted by Pressure CNN (model parameters summarized in \ref{app:xco2CNN}). The yCO$_2$ CNN provides excellent predictions for both the training and validation set (Table~\ref{table}).

\subsubsection{Fluid density CNNs}
The fluid phase densities in the gas or liquid phase depends on the temperature, pressure, and molar fraction of CO$_2$ in that phase. Therefore, we trained two auxiliary CNNs to generate density predictions given temperature, pressure predicted by the Pressure CNN, and molar fraction predicted by the xCO$_2$ or yCO$_2$ CNN (model architecture in \ref{app:xco2CNN}). The trained fluid phase density CNNs are highly accurate (Table~\ref{table}).

\subsubsection{Error analysis}
\begin{figure}[!ht]
\centering
\includegraphics[width=\linewidth]{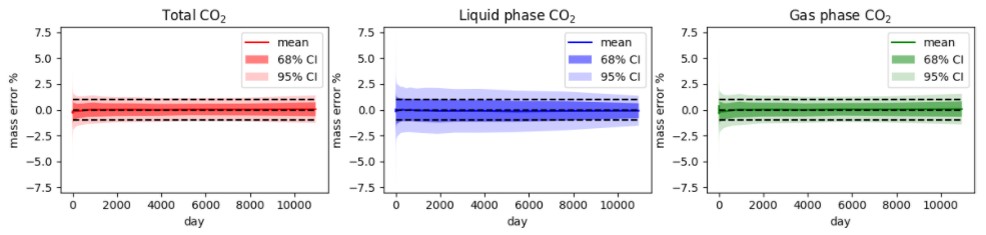}
\caption{Mass balance error for the total, liquid phase, and gas phase CO$_2$ mass. The x-axes indicate the days of injection and the y-axes indicate percentage of the error. The black dotted dash lines are references for $\pm1\%$. The light and dark shaded area are the 68\% and 95\% confidence intervals of the error.}
\label{fig:mass}
\end{figure}

CCSNet generates accurate mass balances over the entire injection period (Table~\ref{table} and Figure~\ref{fig:mass}). The largest errors occur during the first several days of injection. We hypothesize this is caused by a larger fraction of the CO$_2$ being in the liquid phase early in the injection process. The amount of CO$_2$ dissolved in the liquid phase is highly influenced by the artifacts of numerical dispersion at the leading edge of the plume. Therefore, at the beginning of the CO$_2$ injection, the training data are less systematic and challenging to learn. As injection goes on, a larger fraction of the CO$_2$ mass is in the gas phase, therefore the mass predictions becomes more accurate.

Compared to physics informed machine learning approaches, supervised learning methods are criticized for the `lack of physics' because the loss function does not explicitly describe the conservation laws and governing equations. However, our accurate mass balances together with the accurate distribution of CO$_2$ in both phases indicate that the supervised learning-based prediction sequence can satisfy the conservation laws and governing equations given sufficient data and training.

\section{Discussion}
\subsection{Comparative computational efficiency}\label{efficiency}
The average numerical simulation run time for 1,000 random cases using ECLIPSE (e300) is 10 minutes on an Intel\textsuperscript{®} Xeon\textsuperscript{®} Processor E5-2670 CPU. The numerical simulation run time for each simulation varies from 4 to over 100 minutes. Each simulation case utilizes a fully dedicated CPU, and the run time depends on the difficulty of the case.  

To compare the computational efficiency, we used a NVIDIA v100 graphical processing units (GPUs) for CCSNet model inference. CCSNet's prediction times have very small variances ($\sim$1\%) compared to the conventional numerical simulator, and we computed computational efficiency based on the average of 1,000 random samples. The prediction time for the Saturation CNN and Pressure CNN are $\sim$0.05s and $\sim$0.04s, respectively. Given the gas saturation and pressure buildup, the models for predicting the molar fraction of CO$_2$ in the liquid and gas phases each take $\sim$0.03s. The gas and liquid phase fluid densities also require $\sim$0.03s. Therefore, running the entire deep learning model sequence requires $\sim$0.22s to provide the full set of outputs that a numerical simulator can provide. The comparative speed-up between using CCSNet and ECLIPSE varies from $10^{3}$ to $10^{4}$ orders of magnitude depending on the information required by the particular analysis. For example, when predicting the sweep efficiency, we can run the Saturation CNN model by itself, which requires only $\sim$0.05s. Calculating the solubility trapping requires outputs from the Saturation, Pressure, xCO$_2$, and liquid phase density CNNs, which adds up to $\sim$0.15s. Average speed ups for relevant analyses are summarized in Table~\ref{table:computational}. 

\begin{table}[h]
\centering
\footnotesize
\caption{Comparative computational efficiency of CCSNet. We used a NVIDIA v100 GPU for the model inference and the prediction time was calculated by taking the average of 1,000 random runs. The average ECLIPSE simulation run time in the training set (10 mins) was used for the comparison, where each simulation was carried out using an dedicated Intel\textsuperscript{®} Xeon\textsuperscript{®} Processor E5-2670 CPU.}
\begin{tabular}{lll}
Variable &  Prediction time & Average speed up\\
\hline
Gas saturation distribution & 0.05s & 1.2$\times 10^4$ \\
Pressure buildup  & 0.04s & 1.5$\times 10^4$ \\
Sweep efficiency  & 0.05s & 1.2$\times 10^4$ \\
Solubility trapping & 0.15s & 4.0$\times 10^4$ \\
Mass balance & 0.22s & 2.7$\times 10^3$ \\
 \hline
\end{tabular}
\label{table:computational}
\end{table}

\subsection{Applications with fast prediction}
Taking the advantage of the fast prediction speed of CCSNet, we develop a method for estimating sweep efficiency and solubility trapping using information that is often available for screening or comparing different CO$_2$ storage sites. In both cases, we stochastically sample the problem domains to develop a large `data set' composed of 5,000 CCSNet runs to establish a relationship between the reservoir/operational properties and the sweep efficiency or solubility trapping. We sampled the homogeneous permeability from 1000mD to 5mD with a log-uniform distribution and the following variables with an uniform distribution: injection rate from 0.2 to 2 MT/yr, initial pressure from 80 to 160 bar, geothermal gradient from 22 to 28 $^\circ C$/km, $S_{wi}$ from 0.1 to 0.3, $\lambda$ from 0.3 to 0.7, reservoir thickness from 15 to 200m, and perforation length from 15 m to the reservoir thickness. The sampled data set contains only those cases where the maximum pressure buildup is limited to 75\% of the initial reservoir pressure. Here we use homogeneous reservoir characteristics since this is usually the only information available during site screening (e.g. in advance of detailed site-specific studies). By using CCSNet, the computational time for exhaustively sampling the domain is reduced from $\sim$35 days to $\sim$4 mins for sweep efficiency and $\sim$12 mins for solubility trapping.

\subsubsection{Sweep efficiency estimation}
\begin{figure}[h]
\centering
\includegraphics[width=0.5\linewidth]{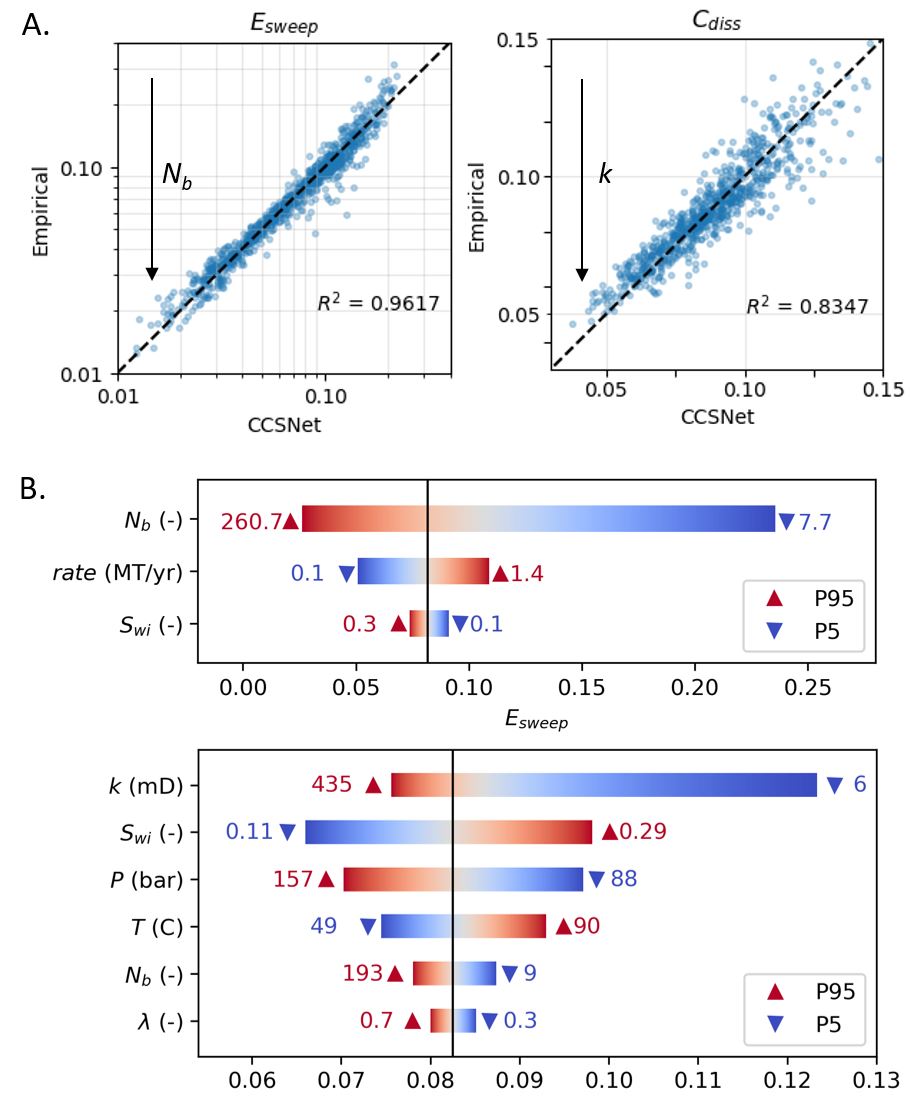}
\caption{A. Comparisons of sweep efficiency and solubility trapping coefficient calculated using the empirical Equation~\ref{eqt:sweep} and~\ref{eqt:solu} verses using CCSNet in the validation set. B. Coefficient sensitivity of each term in Equation~\ref{eqt:sweep} and~\ref{eqt:solu}. P5 and P95 represent the lowest 5$^{th}$ percentile and highest 95$^{th}$ percentile of the value for the term in the validation set. } 
\label{fig:coef}
\end{figure}

Sweep efficiency is a measure of how efficiently the storage space in a reservoir is used; the higher the sweep efficiency the better because higher sweep efficiency results in a smaller footprint of the CO$_2$ plume~\cite{van1995co2}. The footprint is the areal extent of the plume defined as $\pi r_{max}^2$, where $r_{max}$ refers to the largest distance away from the well that CO$_2$ has migrated. Sweep efficiency ($E_{sweep}$) is calculated as: 
\begin{equation}
\begin{split}
E_{sweep}=\frac{V_{gas}}{V_{r_{footprint}}}=\frac{\sum_nV_n\phi_nS_{n}}{\sum_{n\in footprint}V_{n}\phi_{n}},
\label{eqt:sweep_eff} 
\end{split}
\end{equation}
where $V$ is the cell volume, $\phi$ is the porosity, $S$ is the gas saturation, $n$ denotes the spatial grid cell, and $n\in footprint$ denotes all grid cells within the the plume footprint. Using the gas saturation predicted by CCSNet, we find that E$_{sweep}$ depends strongly on reservoir characteristics, ranges from as low as 0.01 to 0.2 over the sampled problem domain (Figure~\ref{fig:coef}A). We use CCSNet-generated $E_{sweep}$  values as inputs into a non-linear multivariate regression algorithm (details in \ref{regression}) to develop an empirical relationship between $E_{sweep}$ and reservoir/operational parameters. 
\begin{equation}
\begin{split}
E_{sweep}=\exp(0.05955-0.5258\ln{N_{b}}-1.390\times 10^{-3}N_{b}  \\ + 0.2503\ln(\frac{r_{inj}}{r_{ref}})-1.162S_{wi}+\epsilon), N_{b}\in (10,450)
\label{eqt:sweep}
\end{split}
\end{equation}

Here $N_{b}$ is the Bond number defined as $\Delta \rho g b_{res}/P_{cap} $ where $b_{res}$ is the reservoir thickness and $P_{cap}$ is the capillary entry pressure, $S_{wi}$ is the irreducible water saturation, $r$ is injection rate, and $\epsilon$ is the error term (reference values and details on each term summarized in Table~\ref{table:fitting}). As shown in Figure~\ref{fig:coef}A, Eq.~\ref{eqt:sweep} is a excellent predictor of sweep efficiency over the range of Bond numbers from 10 to 450, as long as the injection rate is limited to comply with the 75\% overpressure constraint.  Previous studies have demonstrated that sweep efficiency and trapping are influenced by gravity number and depositional environment~\cite{Ide2007, osti_1177419}. Here we show that for homogeneous reservoirs, the Bond number ($N_b$) has the largest influence on $E_{sweep}$; reservoirs with lower Bond numbers have higher $E_{sweep}$ (Figure~\ref{fig:coef}B). $E_{sweep}$ also increases with higher injection rates and lower S$_{wi}$ values. Surprisingly, for the homogeneous reservoirs studied here, factors such as injection depth or injection interval had no significant influence on $E_{sweep}$. This would be expected to change for heterogeneous reservoirs.

\subsubsection{Solubility trapping estimation}
Solubility trapping occurs when CO$_2$ dissolves into the formation water and is beneficial for reducing the risk of CO$_2$ leakage~\cite{gunter2004role, gilfillan2009solubility, suekane2008geological}. Using a similar stochastic approach as described above, we developed an empirical expression for estimating solubility trapping ($C_{diss}$), where $C_{diss}$ is the mass fraction of the injected CO$_2$ dissolved in the formation water:
\begin{equation}
\begin{aligned}
C_{diss}= &0.0762 + 0.1804S_{wi} - 0.0030\ln{N_b}+ 0.2667\frac{k_{ref}}{k}\\
        &- 0.0149\lambda - 0.0964\frac{P}{P_{ref}} + 0.0177\frac{T}{T_{ref}}+\epsilon
\label{eqt:solu}
\end{aligned}
\end{equation}
where $k$ is permeability, $\lambda$ is the coefficient in van Genechuten capillary function, $P$ is the initial pressure, and $T$ is the temperature. Reference values and the error term are summarized in \ref{table:fitting}. Solubility trapping is strongly influenced by a number of reservoir properties, with values ranging from 0.05 to nearly 0.15 (Figure~\ref{fig:coef}A). In addition to the Bond number and $S_{wi}$, $C_{diss}$ is also influenced by the formation permeability, initial pressure, temperature, and the coefficient in the capillary pressure function. Solubility trapping increases with lower permeability, higher reservoir temperature, lower pressure, and lower Bond number (Figure~\ref{fig:coef}B). This result is counterintuitive, because the solubility of CO$_2$ in water increases with higher pressure and lower temperature. This analysis suggests $C_{diss}$ is more strongly controlled by the density of the CO$_2$, which like $C_{diss}$, decreases with higher temperature and lower pressure. The lower the density of CO$_2$, the greater the plume volume, hence more contact area with the formation water and higher $C_{diss}$.

Note that the solubility trapping described here occurs during the injection phase of the CO$_2$ storage project. After injection stops, CO$_2$ will continue to dissolve as the result of convective mixing~\cite{riaz2006onset} and spreading~\cite{macminn2012spreading}, thus our estimates should be viewed as a lower bound. Additionally, we considered reservoirs with pure water; reservoirs with higher concentrations of dissolved salts have lower solubility~\cite{enick1990co2} and consequently smaller amounts of solubility trapping. 

\section{Conclusion and future work}

We show that deep learning models such as CCSNet can provide an alternative to computationally intensive numerical simulators for routine tasks, such as predicting the injection performance of CO$_2$ storage projects. Important parameters such as the maximum extent of the CO$_2$ plume, saturation distributions, pressure buildup at the injection well and throughout the reservoir, sweep efficiency, and solubility trapping can be calculated accurately with high computational efficiency. While CCSNet includes many of the important parameters needed to realistically simulate the injection phase of a CO$_2$ storage projects, it is currently limited to systems well-represented by 2d-radial geometry and isotropic rock properties, and does not yet include post-injection processes processes such as residual gas trapping~\cite{doughty2007modeling, krevor2015capillary} or mineral trapping~\cite{bachu1994aquifer, xu2004numerical}. Nevertheless, the ability to train the model to perform the current tasks with such a high degree of accuracy, covering such a large domain of input parameters bodes well for increasing the capabilities of these models to include other features.

CCSNet has the flexibility to include additional input parameters and features when needed. As discussed in data configuration, the input volumes contain idle slices that can be used for other parameters. For example, the model currently uses isotropic permeability values. We can easily add anisotropy and porosity to the model by converting some slices into directional permeability and porosity. Similarly, relative permeability curves could be modified to include residual gas trapping~\cite{doughty2007modeling}.

The benefits of the high computational efficiency of CCSNet are evident from the new methods for estimating $E_{sweep}$ and $C_{diss}$ presented here. It is now possible to quickly provide reservoir-specific estimates of these parameters for screening prospective storage sites using data sets that are publicly available~\cite{doecarbon, USGS, BOEM}. 
CCSNet can also be used once more site-specific data on geological heterogeneity is available to optimize injection depths and rates, make proabalistic predictions of plume footprint and pressure buildup, and for inverse modeling of monitoring data; all tasks required to support regulatory permit applications and compliance.

\section*{Web application}
We developed a publicly accessible web application that hosts CCSNet. Users' can customize input variable combinations, including uploading their own permeability maps. The web application provides both independent and collaborative prediction to the models described above and produces outputs such as gas saturation, pressure buildup, solubility trapping, and sweep efficiency. Refer to \url{https://youtu.be/5bIlfjyo6Jk} for a video demonstration of this web application. The web application will be released to public upon the publication of this manuscript. 

\section*{Code and data availability}
The python code for CCSNet modeling suite and the data set used in training will be released upon the publication of this manuscript.

\section*{Acknowledgments}
This work was supported by ExxonMobil through the Strategic Energy Alliance at Stanford University and the Stanford Center for Carbon Storage.

\appendix

\newpage
\section{Grid resolution}
\label{GridResolution}
The vertical grid dimension is $d_z = 200m/96 = 2.0833m$. The radial grid dimension is $d_r = 3.6m \times a^{i-1}$, where $a = 1.035012, i \in [1,...,200]$. We used 24 time snapshots with gradually coarsening resolution to represent the total 30-year period where the time interval varies from days to years. Time step intervals $d_t = 1.421245^{i-1}$ days, where $i \in [1,...,24]$. At early time steps, the CO$_2$ plume is located near the injection well where the spatial grid has high resolution. Capturing the variability between each time step requires high temporal resolution. Towards the end of the injection, because the plume migrates away from the injection well where the spatial grids are coarser, the coarse time resolutions are adequate. The temporal grid design also satisfies needs for operation that often requires finer time resolution monitoring at the beginning of injection. 

\newpage
\section{Statistical characteristics of permeability maps}\label{Statistical}
\begin{table}[!ht]
\centering
\scriptsize
\caption{The permeability maps are generated using Stanford Geostatistical Modeling Software (SGeMS)~\cite{remy_boucher_wu_2009}. SGeMS is an open-source computer package for geostatistical modeling according to user defined spatial variables. Here we defined the medium appearance, spatial correlation, mean, and contrast ratio ($k_{high}/k_{low}$) in each map to create a large variety of permeability maps. The permeability value for an individual cell can range from 10$^{-3}$ mD to 10$^2$ D.}
\begin{tabular}{lllllll}
    \hline
    Medium & Parameter & Mean & Std & Max & Min & Unit\\
    \hline
    Gaussian & Field average & 30.8 & 58.3 & 1053 & 0.3 & mD \\
             & Vertical correlation & 7.3 & 3.6 & 12.5 & 2.1 & m \\
             & Horizontal correlation & 2190 & 1432 & 6250 & 208 & m \\
             & Contrast ratio & 4.01$\times$ 10$^4$ & 2.19$\times$ 10$^5$ & 3.00$\times$ 10$^6$ & 1.01 & - \\
  \hline
    von Karman~\cite{carpentier2009conservation} & Field average & 39.9 & 54.4 & 867.9 & 1.8 & mD\\
             & Vertical correlation & 7.2 & 3.5 & 12.5 & 2.1 & m\\
             & Horizontal correlation & 2.15$\times$ 10$^4$ & 1.40$\times$ 10$^4$ & 6.23$\times$ 10$^4$ & 208 & m\\             
             & Contrast ratio & 2.66$\times$ 10$^4$ & 1.54$\times$ 10$^5$ &  2.12$\times$ 10$^6$ & 1.00 & - \\     
  \hline
    Discontinuous & Field average & 80.8 & 260.2 & 5281 & 2.0 & mD\\
             & Vertical correlation & 7.2 & 3.6 & 12.5 & 2.1 & m\\
             & Horizontal correlation & 2176 & 1429 & 6250 & 208 & m\\             
             & Contrast ratio & 2.17$\times$ 10$^4$ & 1.51$\times$ 10$^5$ & 2.68$\times$ 10$^6$ & 1.01 & - \\               
  \hline
    Layered  & Field average & 258.6 & 140.8 & 1022 & 5.4 & mD\\
             & Number of materials & 10 & 5 & 20 & 2 & -\\             
             & Contrast ratio & 190.7 & 582.0 & 1.38$\times$ 10$^4$ & 1.00 & - \\                    
  \hline
    Homogeneous & Field permeability & 327.7 & 478.1 & 1216 & 4.0 & mD\\
  \hline
  \end{tabular}
\label{sgems}
\end{table}

\newpage
\section{Mass balance analysis}
\label{MassBalance}
The discretized form of the mass accumulation term (Equation~\ref{mass_balance}) at a given time step is written as:
\begin{equation}
M=\sum_{i,n}V_n(\phi S\rho X)_{i,n},
\end{equation}
where and $n$ denotes the spatial grid and $i$ denotes phase $gas$ or $liquid$. CCSNet uses six models to collaboratively provide predictions to all the components. The Saturation CNN predicts the gas saturation which provides $S_{i,n}$; the fluid density CNNs provide $\rho_{i,n}$; the xCO$_2$ CNN and the yCO$_2$ CNN provide the molar fraction of CO$_2$ in each phase, which are thus used for calculating $X_{i,n}$. Since the reservoir rock is compressible, pore volume $\phi_n$ is a function of the pressure in each cell:
\begin{equation}
C_t = \frac{d\phi_n}{dP_n}\frac{1}{\phi_n},
\end{equation}
where $C_t=5\times 10^{-4}$ bar$^{-1}$ represents the rock compressibility. The Pressure CNN predicts the pressure in each grid cell and the pore volume $\phi_n$ is adjusted as $\phi_n(P_n) =\phi_n(P_{n,ref})(1+X+{X^2}/{2})$, where $X = C_t(P_{n}-P_{n,ref})$ and $P_{n,ref} = 1.0132$ bar. 

Note that a neural network rarely predicts true zeros because the outputs are calculated empirically. Instead, zeros are represented by tiny numbers such as $10^{-6}$. In the mass balance calculation, these tiny values in the predictions are amplified at grid cells that are far away from the injection well due to the large grid cell volume. Therefore, we applied cutoffs to the prediction of the Saturation and xCO$_2$ CNNs by casting gas saturation smaller than 1e-2 and xCO$_2$ smaller than 8e-4 to be zeros.

\newpage
\section{Saturation CNN architecture}\label{app:SaturationCNN}
\begin{table}[!ht]
  \centering
  \footnotesize
  \caption{Saturation CNN architecture. \texttt{Conv3D} denotes a 3D convolutional layer; \texttt{c} denotes the number of channels in the layer output; \texttt{k} denotes the kernel (also refered as filter) size; \texttt{s} denotes the size of the stride; \texttt{BN} denotes a batch normalization layer; \texttt{ReLu} denotes a rectified linear layer, \texttt{Add} denotes a addition with the identity; \texttt{UnSampling} denotes a unSampling layer that expands the matrix dimension using nearest neighbor method, and \texttt{Padding} denotes a padding layer using the reflection padding technique. In this model, the number of total parameters is 40,399,489 with 40,386,817 trainable parameters and 12,672 non-trainable parameters.}
  \begin{tabular}{lll}
    \hline
    Part     & Layer     & Output Shape \\
    \hline
    Input    &                                   & (96,200,24,1) \\
    Encode 1 &  \texttt{Conv3D(c32k3s2)/BN/ReLu} & (48,100,12,32)  \\
    Encode 2 &  \texttt{Conv3D(c64k3s1)/BN/ReLu} & (48,100,12,64)  \\
    Encode 3 &  \texttt{Conv3D(c128k3s2)/BN/ReLu} & (24,50,6,128)  \\
    Encode 4 &  \texttt{Conv3D(c128k3s1)/BN/ReLu} & (24,50,6,128)  \\
    Encode 5 &  \texttt{Conv3D(c256k3s2)/BN/ReLu} & (12,25,3,256)  \\
    Encode 6 &  \texttt{Conv3D(c256k3s1)/BN/ReLu} & (12,25,3,256)  \\
   ResConv 1 &  \texttt{Conv3D(c256k3s1)/BN/Conv3D(c256k3s1)/BN/ReLu/Add} &  (12,25,3,256)  \\
   ResConv 2 &  \texttt{Conv3D(c256k3s1)/BN/Conv3D(c256k3s1)/BN/ReLu/Add} &  (12,25,3,256)  \\
   ResConv 3 &  \texttt{Conv3D(c256k3s1)/BN/Conv3D(c256k3s1)/BN/ReLu/Add} &  (12,25,3,256)  \\
   ResConv 4 &  \texttt{Conv3D(c256k3s1)/BN/Conv3D(c256k3s1)/BN/ReLu/Add} &  (12,25,3,256)  \\
   ResConv 5 &  \texttt{Conv3D(c256k3s1)/BN/Conv3D(c256k3s1)/BN/ReLu/Add} &  (12,25,3,256)  \\
   ResConv 6 &  \texttt{Conv3D(c256k3s1)/BN/Conv3D(c256k3s1)/BN/ReLu/Add} &  (12,25,3,256)  \\
   ResConv 7 &  \texttt{Conv3D(c256k3s1)/BN/Conv3D(c256k3s1)/BN/ReLu/Add} &  (12,25,3,256)  \\
   ResConv 8 &  \texttt{Conv3D(c256k3s1)/BN/Conv3D(c256k3s1)/BN/ReLu/Add} &  (12,25,3,256)  \\
    Decode 6 &  \texttt{UnSampling/Padding/Conv3D(c256k3s1)/BN/Relu} &  (12,25,3,256) \\
    Decode 5 &  \texttt{UnSampling/Padding/Conv3D(c256k3s2)/BN/Relu} & (24,50,6,256)  \\
    Decode 4 &  \texttt{UnSampling/Padding/Conv3D(c128k3s1)/BN/Relu} & (24,50,6,128)  \\
    Decode 3 &  \texttt{UnSampling/Padding/Conv3D(c128k3s2)/BN/Relu} & (48,100,12,128)  \\
    Decode 2 &  \texttt{UnSampling/Padding/Conv3D(c64k3s1)/BN/Relu} & (48,100,12,64)  \\
    Decode 1 &  \texttt{UnSampling/Padding/Conv3D(c32k3s2)/BN/Relu} & (96,200,24,32)\\
    Output   &  \texttt{Conv3D(c1k3s1)}                         & (96,200,24,1) \\
    \hline
  \end{tabular}
  \label{table:saturationcnn}
\end{table}

\newpage
\section{Pressure CNN architecture}\label{app:PressureCNN}
\begin{table}[!ht]
\centering
\footnotesize
\caption{Pressure CNN architecture. \texttt{Conv3D} denotes a 3D convolutional layer; \texttt{c} denotes the number of channels in the layer output; \texttt{k} denotes the kernel (also refered as filter) size; \texttt{s} denotes the size of the stride; \texttt{BN} denotes a batch normalization layer; \texttt{ReLu} denotes a rectified linear layer, \texttt{Add} denotes a addition with the identity; \texttt{UnSampling} denotes a UnSampling layer that expands the matrix dimension using nearest neighbor method, and \texttt{Padding} denotes a padding layer using the reflection padding technique. Total params: 33,316,481, trainable params: 33,305,857, non-trainable params: 10,624.
}
  \begin{tabular}{lll}
    \hline
    Part     & Layer     & Output Shape \\
    \hline
    Input    &                                   & (96,200,24,1) \\
    Encode 1 &  \texttt{Conv3D(c32k3s2)/BN/ReLu} & (48,100,12,32)  \\
    Encode 2 &  \texttt{Conv3D(c64k3s1)/BN/ReLu} & (48,100,12,64)  \\
    Encode 3 &  \texttt{Conv3D(c128k3s2)/BN/ReLu} & (24,50,6,128)  \\
    Encode 4 &  \texttt{Conv3D(c128k3s1)/BN/ReLu} & (24,50,6,128)  \\
    Encode 5 &  \texttt{Conv3D(c256k3s2)/BN/ReLu} & (12,25,3,256)  \\
    Encode 6 &  \texttt{Conv3D(c256k3s1)/BN/ReLu} & (12,25,3,256)  \\
   ResConv 1 &  \texttt{Conv3D(c256k3s1)/BN/Conv3D(c256k3s1)/BN/ReLu/Add} &  (12,25,3,256)  \\
   ResConv 2 &  \texttt{Conv3D(c256k3s1)/BN/Conv3D(c256k3s1)/BN/ReLu/Add} &  (12,25,3,256)  \\
   ResConv 3 &  \texttt{Conv3D(c256k3s1)/BN/Conv3D(c256k3s1)/BN/ReLu/Add} &  (12,25,3,256)  \\
   ResConv 4 &  \texttt{Conv3D(c256k3s1)/BN/Conv3D(c256k3s1)/BN/ReLu/Add} &  (12,25,3,256)  \\
   ResConv 5 &  \texttt{Conv3D(c256k3s1)/BN/Conv3D(c256k3s1)/BN/ReLu/Add} &  (12,25,3,256)  \\
   ResConv 6 &  \texttt{Conv3D(c256k3s1)/BN/Conv3D(c256k3s1)/BN/ReLu/Add} &  (12,25,3,256)  \\
    Decode 6 &  \texttt{UnSampling/Padding/Conv3D(c256k3s1)/BN/Relu} &  (12,25,3,256) \\
    Decode 5 &  \texttt{UnSampling/Padding/Conv3D(c256k3s2)/BN/Relu} & (24,50,6,256)  \\
    Decode 4 &  \texttt{UnSampling/Padding/Conv3D(c128k3s1)/BN/Relu} & (24,50,6,128)  \\
    Decode 3 &  \texttt{UnSampling/Padding/Conv3D(c128k3s2)/BN/Relu} & (48,100,12,128)  \\
    Decode 2 &  \texttt{UnSampling/Padding/Conv3D(c64k3s1)/BN/Relu} & (48,100,12,64)  \\
    Decode 1 &  \texttt{UnSampling/Padding/Conv3D(c32k3s2)/BN/Relu} & (96,200,24,32)\\
    Output   &  \texttt{Conv3D(c1k3s1)}                         & (96,200,24,1) \\
    \hline
  \end{tabular}
  \label{pressurecnn}
\end{table}

\newpage
\section{xCO$_2$, yCO$_2$, and fluid densities CNN architecture}\label{app:xco2CNN}

\begin{table}[!ht]
  \centering
  \footnotesize
  \caption{xCO$_2$, yCO$_2$, gas phase density, and liquid phase density CNN architecture. \texttt{Conv3D} denotes a 3D convolutional layer; \texttt{c} denotes the number of channels in the layer output; \texttt{k} denotes the kernel (also referred to as filter) size; \texttt{s} denotes the size of the stride; \texttt{BN} denotes a batch normalization layer; \texttt{ReLu} denotes an rectified linear activation layer, \texttt{Add} denotes a addition with the identity; \texttt{UnSampling} denotes a UnSampling layer that expands the matrix dimension using nearest neighbor method, and \texttt{Padding} denotes a padding layer using the reflection padding technique. Total params: 8,337,057, trainable params: 8,331,745, non-trainable params: 5,312.}
  \begin{tabular}{lll}
    \hline
    Part     & Layer     & Output Shape \\
    \hline
    Input    &                                   & (96,200,24,3) \\
    Encode 1 &  \texttt{Conv3D(c16k3s2)/BN/ReLu} & (48,100,12,16)  \\
    Encode 2 &  \texttt{Conv3D(c32k3s1)/BN/ReLu} & (48,100,12,32)  \\
    Encode 3 &  \texttt{Conv3D(c64k3s2)/BN/ReLu} & (24,50,6,64)  \\
    Encode 4 &  \texttt{Conv3D(c64k3s1)/BN/ReLu} & (24,50,6,64)  \\
    Encode 5 &  \texttt{Conv3D(c128k3s2)/BN/ReLu} & (12,25,3,128)  \\
    Encode 6 &  \texttt{Conv3D(c128k3s1)/BN/ReLu} & (12,25,3,128)  \\
   ResConv 1 &  \texttt{Conv3D(c128k3s1)/BN/Conv3D(c128k3s1)/BN/ReLu/Add} &  (12,25,3,128)  \\
   ResConv 2 &  \texttt{Conv3D(c128k3s1)/BN/Conv3D(c128k3s1)/BN/ReLu/Add} &  (12,25,3,128)  \\
   ResConv 3 &  \texttt{Conv3D(c128k3s1)/BN/Conv3D(c128k3s1)/BN/ReLu/Add} &  (12,25,3,128)  \\
   ResConv 4 &  \texttt{Conv3D(c128k3s1)/BN/Conv3D(c128k3s1)/BN/ReLu/Add} &  (12,25,3,128)  \\
   ResConv 5 &  \texttt{Conv3D(c128k3s1)/BN/Conv3D(c128k3s1)/BN/ReLu/Add} &  (12,25,3,128)  \\
   ResConv 6 &  \texttt{Conv3D(c128k3s1)/BN/Conv3D(c128k3s1)/BN/ReLu/Add} &  (12,25,3,128)  \\
   ResConv 7 &  \texttt{Conv3D(c128k3s1)/BN/Conv3D(c128k3s1)/BN/ReLu/Add} &  (12,25,3,128)  \\
    Decode 6 &  \texttt{Unpool/Padding/Conv3D(c128k3s1)/BN/Relu} &  (12,25,3,128) \\
    Decode 5 &  \texttt{Unpool/Padding/Conv3D(c128k3s2)/BN/Relu} & (24,50,6,128)  \\
    Decode 4 &  \texttt{Unpool/Padding/Conv3D(c64k3s1)/BN/Relu} & (24,50,6,64)  \\
    Decode 3 &  \texttt{Unpool/Padding/Conv3D(c64k3s2)/BN/Relu} & (48,100,12,64)  \\
    Decode 2 &  \texttt{Unpool/Padding/Conv3D(c32k3s1)/BN/Relu} & (48,100,12,32)  \\
    Decode 1 & \texttt{Unpool/Padding/Conv3D(c16k3s2)/BN/Relu} & (96,200,24,16)\\
    Output   &  \texttt{Conv3D(c1k3s1)}                         & (96,200,24,1) \\
    \hline
  \end{tabular}
  \label{xco2cnn}
\end{table}

\newpage
\section{Multivariate regression}\label{regression}
Using the data set described above, we randomly split 80\% of the data into the training set and 20\% into the validation set to develop the relationship between sweep efficiency, solubility trapping and a large variety of dimensionless variables. The variables include dimensionless numbers such as Bond Number ($N_{b}= \Delta \rho g b_{res}/P_{c}$ where $b_{res}$ is reservoir thickness), Capillary number ($N_c$), and Gravity number ($N_g$), as well as dimensionless reservoir properties, including permeability (${k}/{k_{ref}}$), initial pressure (${P}/{P_{ref}}$), injection rate (${r_{inj}}/{r_{ref}}$), perforation thickness to reservoir thickness (${b_{perf}}/{b_{res}}$ where $b_{perf}$ is perforation thickness), perforation depth to reservoir thickness (${l_{perf}}/{b_{res}}$ where $l_{perf}$ is perforation depth from the reservior top), irreducible water saturation ($S_{wi}$), and capillary pressure curve scaling factor ($\lambda$ in Equation~\ref{eqs:lambda}). We also investigated various combinations and variations (such as reciprocals) of these aforementioned variables.  

Using those relationships to inform our model, we ran a sequence of single-variable and multi-variable linear and nonlinear regressions on a training data set. We used forward variable selection with criteria of $R^2$, Adjusted $R^2$, and root mean squared error (RMSE) to assess the quality of the model. The Adjusted $R^2$ penalizes additional variables to reduce the number of variables used in the prediction. Concurrently, we used the Normal Q-Q plot to examine whether the residuals were normally distributed; the scale-location plot to monitor the constant variance assumption; the residuals versus fitted values plot to evaluate whether the data set shows non-constant variance or non-linear trends; and the Cook's distance plots to to identify outliers that might significantly influence the model. 

Using these diagnostic plots and the quality criterion, we discovered the optimal model for sweep efficiency with $N_b$, ${r_{inj}}/{r_{ref}}$, and $S_{wi}$, and for solubility trapping, a model with $S_{wi}$, $N_b$, $\lambda$, ${k_{ref}}/{k}$, ${P}/{P_{ref}}$, and ${T}/{T_{ref}}$. Details on the models' parameters and criterion results are summarized in Table \ref{table:fitting}.

\begin{table}
\centering
  \scriptsize
\caption{Prediction ranges, constants, quality criterion, and term standard errors for the sweep efficiency and solubility trapping estimation equations. Standard error for each term is calculated based on the the term value and the training set.}
\begin{tabular}{l|l|llll}
    \hline
    Sweep efficiency    & category & parameter & value & unit  \\
    \hline
 & Prediction range & $N_b$ & 10 to  450  & -  \\ 
                 &  & $r_{inj}$ & 0.02 to 2.0 & MT/yr \\
                 &  & $S_{wi}$ & 0.1 to 0.3 & -  \\
    \hline
                  & Constant & $r_{ref}$ & 1 & MT/yr \\
                  &  & $\epsilon$ & 0.01266 & - \\
    \hline
                 & Standard error & $ln{(N_b)}$ & 3.56e-03 & - \\
                 & &  $N_b$ & 4.38e-05 & -\\
                 & & $ln{(\frac{r_{inj}}{r_{ref}})}$  & 2.32e-03 & - \\
                 & &  $S_{wi}$& 3.02e-02 & - \\
     \hline
                 & Quality criterion & training RMSE & 0.109 & - & \\
                 & & training  R-Squared & 0.964 & - & \\
                 & & training Adjusted R-Squared & 0.964 & - & \\
                 & & validation RMSE & 0.116 & - & \\
                 & & validation R-Squared & 0.962 & - & \\
                 & & validation Adjusted R-Squared & 0.962 & - & \\
     \hline
     \hline
Solubility trapping   & category & parameter & value & unit  \\
     \hline
 & Prediction range & $S_{wi}$ & 0.1 to 0.3 & -  \\
                     &  & $N_b$ & 10 to  450  & -  \\ 
                     &  & $k$ & 5 to  1000  & mD  \\ 
                     &  & $\lambda$ & 0.3 to  0.7  & -  \\ 
                     &  & $P$ & 80 to 160  & bar  \\ 
                     &  & $T$ & 40 to 100  & C$^\circ$  \\ 
     \hline
                  & Constant & $k_{ref}$ & 1 & mD \\
                  &          & $P_{ref}$ & 250 & bar \\
                  &          & $T_{ref}$ & 40 & C$^\circ$ \\
                  &  & $\epsilon$ & 6.0165e-5 & - \\
    \hline
  & Standard error & $ln{(N_b)}$ & 2.10e-04 & -\\
  &  & $\frac{k_{ref}}{k}$ & 3.94e-03 & -\\
  &  & $\lambda$ & 1.28e-03 & -\\
  &  & $S_{wi}$ & 2.42e-03 & -\\
  &  & $\frac{T}{T_{ref}}$ & 6.30e-04 & - \\
  &  & $\frac{P}{P_{ref}}$ &  2.27e-03 & -\\
  \hline
 & Quality criterion & training RMSE & 0.008 & -  \\
 & & training  R-Squared & 0.845 & -  \\
 & & training Adjusted R-Squared & 0.845 & -  \\
 & & validation RMSE & 0.008 & -  \\
 & & validation  R-Squared & 0.833 & -  \\
 & & validation Adjusted R-Squared & 0.834 & -  \\
    \hline
  \end{tabular}
  \label{table:fitting}
\end{table}

\newpage
\bibliographystyle{elsarticle-num} 
\bibliography{ref}

\begin{thebibliography}{10}
\expandafter\ifx\csname url\endcsname\relax
  \def\url#1{\texttt{#1}}\fi
\expandafter\ifx\csname urlprefix\endcsname\relax\def\urlprefix{URL }\fi
\expandafter\ifx\csname href\endcsname\relax
  \def\href#1#2{#2} \def\path#1{#1}\fi

\bibitem{aziz1979petroleum}
K.~Aziz, Petroleum reservoir simulation, Applied Science Publishers 476 (1979).

\bibitem{pachauri2014climate}
R.~K. Pachauri, M.~R. Allen, V.~R. Barros, J.~Broome, W.~Cramer, R.~Christ,
  J.~A. Church, L.~Clarke, Q.~Dahe, P.~Dasgupta, et~al., Climate change 2014:
  synthesis report. Contribution of Working Groups I, II and III to the fifth
  assessment report of the Intergovernmental Panel on Climate Change, Ipcc,
  2014.

\bibitem{allen1985numerical}
M.~B. Allen~III, Numerical modelling of multiphase flow in porous media,
  Advances in Water Resources 8~(4) (1985) 162--187.

\bibitem{chierici1995simulation}
G.~L. Chierici, The simulation of reservoir behaviour using numerical
  modelling, in: Principles of Petroleum Reservoir Engineering, Springer, 1995,
  pp. 123--229.

\bibitem{pruess2005eco2n}
K.~Pruess, ECO2N: A TOUGH2 fluid property module for mixtures of water, NaCl,
  and CO2, Lawrence Berkeley National Laboratory Berkeley, CA, 2005.

\bibitem{khebzegga2020continuous}
O.~Khebzegga, A.~Iranshahr, H.~Tchelepi, Continuous relative permeability model
  for compositional simulation, Transport in Porous Media 134~(1) (2020)
  139--172.

\bibitem{orr2007theory}
F.~M. Orr, et~al., Theory of gas injection processes, Vol.~5, Tie-Line
  Publications Copenhagen, 2007.

\bibitem{pini2012capillary}
R.~Pini, S.~C. Krevor, S.~M. Benson, Capillary pressure and heterogeneity for
  the co2/water system in sandstone rocks at reservoir conditions, Advances in
  Water Resources 38 (2012) 48--59.

\bibitem{Doughty2010}
C.~Doughty, Investigation of co 2 plume behavior for a large-scale pilot test
  of geologic carbon storage in a saline formation, Transport in porous media
  82~(1) (2010) 49--76.

\bibitem{Wen2019}
G.~Wen, S.~M. Benson,
  \href{https://linkinghub.elsevier.com/retrieve/pii/S1750583619300246}{{CO2
  plume migration and dissolution in layered reservoirs}}, International
  Journal of Greenhouse Gas Control 87~(May) (2019) 66--79.
\newblock \href {https://doi.org/10.1016/j.ijggc.2019.05.012}
  {\path{doi:10.1016/j.ijggc.2019.05.012}}.
\newline\urlprefix\url{https://linkinghub.elsevier.com/retrieve/pii/S1750583619300246}

\bibitem{Kitanidis2015}
P.~K. Kitanidis, {Persistent questions of heterogeneity, uncertainty, and scale
  in subsurface flow and transport}, Water Resources Research 51~(8) (2015)
  5888--5904.
\newblock \href {https://doi.org/10.1002/2015WR017639}
  {\path{doi:10.1002/2015WR017639}}.

\bibitem{cardoso2009development}
M.~A. Cardoso, L.~J. Durlofsky, P.~Sarma, Development and application of
  reduced-order modeling procedures for subsurface flow simulation,
  International journal for numerical methods in engineering 77~(9) (2009)
  1322--1350.

\bibitem{razavi2012review}
S.~Razavi, B.~A. Tolson, D.~H. Burn, Review of surrogate modeling in water
  resources, Water Resources Research 48~(7) (2012).

\bibitem{bazargan2015surrogate}
H.~Bazargan, M.~Christie, A.~H. Elsheikh, M.~Ahmadi, Surrogate accelerated
  sampling of reservoir models with complex structures using sparse polynomial
  chaos expansion, Advances in Water Resources 86 (2015) 385--399.

\bibitem{hamdi2017gaussian}
H.~Hamdi, I.~Couckuyt, M.~C. Sousa, T.~Dhaene, Gaussian processes for
  history-matching: application to an unconventional gas reservoir,
  Computational Geosciences 21~(2) (2017) 267--287.

\bibitem{tian2017gaussian}
L.~Tian, R.~Wilkinson, Z.~Yang, H.~Power, F.~Fagerlund, A.~Niemi, Gaussian
  process emulators for quantifying uncertainty in co2 spreading predictions in
  heterogeneous media, Computers \& Geosciences 105 (2017) 113--119.

\bibitem{tang2020deep}
M.~Tang, Y.~Liu, L.~J. Durlofsky, A deep-learning-based surrogate model for
  data assimilation in dynamic subsurface flow problems, Journal of
  Computational Physics 413 (2020) 109456.

\bibitem{jin2020deep}
Z.~L. Jin, Y.~Liu, L.~J. Durlofsky, Deep-learning-based surrogate model for
  reservoir simulation with time-varying well controls, Journal of Petroleum
  Science and Engineering 192 (2020) 107273.

\bibitem{mo2019deep}
S.~Mo, Y.~Zhu, N.~Zabaras, X.~Shi, J.~Wu, Deep convolutional encoder-decoder
  networks for uncertainty quantification of dynamic multiphase flow in
  heterogeneous media, Water Resources Research 55~(1) (2019) 703--728.

\bibitem{zhong2021deep}
Z.~Zhong, A.~Y. Sun, B.~Ren, Y.~Wang, A deep-learning-based approach for
  reservoir production forecast under uncertainty, SPE Journal 1  1--27.

\bibitem{IEA2020ccus}
P.~IEA,
  \href{https://www.iea.org/reports/ccus-in-clean-energy-transitions}{Ccus in
  clean energy transitions}, Tech. rep. (2020).
\newline\urlprefix\url{https://www.iea.org/reports/ccus-in-clean-energy-transitions}

\bibitem{yamamoto2011investigation}
H.~Yamamoto, C.~Doughty, Investigation of gridding effects for numerical
  simulations of co2 geologic sequestration, International Journal of
  Greenhouse Gas Control 5~(4) (2011) 975--985.

\bibitem{Yamamoto2011}
H.~Yamamoto, C.~Doughty,
  \href{http://dx.doi.org/10.1016/j.ijggc.2011.02.007}{{Investigation of
  gridding effects for numerical simulations of CO2 geologic sequestration}},
  International Journal of Greenhouse Gas Control 5~(4) (2011) 975--985.
\newblock \href {https://doi.org/10.1016/j.ijggc.2011.02.007}
  {\path{doi:10.1016/j.ijggc.2011.02.007}}.
\newline\urlprefix\url{http://dx.doi.org/10.1016/j.ijggc.2011.02.007}

\bibitem{Saadatpoor2010}
E.~Saadatpoor, S.~L. Bryant, K.~Sepehrnoori, {New trapping mechanism in carbon
  sequestration}, Transport in Porous Media 82~(1) (2010) 3--17.
\newblock \href {https://doi.org/10.1007/s11242-009-9446-6}
  {\path{doi:10.1007/s11242-009-9446-6}}.

\bibitem{Krevor2015}
S.~Krevor, M.~J. Blunt, S.~M. Benson, C.~H. Pentland, C.~Reynolds,
  A.~Al-Menhali, B.~Niu,
  \href{http://dx.doi.org/10.1016/j.ijggc.2015.04.006}{{Capillary trapping for
  geologic carbon dioxide storage - From pore scale physics to field scale
  implications}}, International Journal of Greenhouse Gas Control 40 (2015)
  221--237.
\newblock \href {https://doi.org/10.1016/j.ijggc.2015.04.006}
  {\path{doi:10.1016/j.ijggc.2015.04.006}}.
\newline\urlprefix\url{http://dx.doi.org/10.1016/j.ijggc.2015.04.006}

\bibitem{Ide2007}
S.~T. Ide, K.~Jessen, F.~M. Orr, {Storage of CO2 in saline aquifers: Effects of
  gravity, viscous, and capillary forces on amount and timing of trapping},
  International Journal of Greenhouse Gas Control 1~(4) (2007) 481--491.
\newblock \href {https://doi.org/10.1016/S1750-5836(07)00091-6}
  {\path{doi:10.1016/S1750-5836(07)00091-6}}.

\bibitem{Pruess2011}
K.~Pruess, J.~Nordbotten, Numerical simulation studies of the long-term
  evolution of a co 2 plume in a saline aquifer with a sloping caprock,
  Transport in porous media 90~(1) (2011) 135--151.

\bibitem{Zhu2019}
Y.~Zhu, N.~Zabaras, P.~S. Koutsourelakis, P.~Perdikaris,
  \href{https://doi.org/10.1016/j.jcp.2019.05.024}{{Physics-constrained deep
  learning for high-dimensional surrogate modeling and uncertainty
  quantification without labeled data}}, Journal of Computational Physics 394
  (2019) 56--81.
\newblock \href {http://arxiv.org/abs/1901.06314} {\path{arXiv:1901.06314}},
  \href {https://doi.org/10.1016/j.jcp.2019.05.024}
  {\path{doi:10.1016/j.jcp.2019.05.024}}.
\newline\urlprefix\url{https://doi.org/10.1016/j.jcp.2019.05.024}

\bibitem{fuks2020physics}
O.~Fuks, H.~Tchelepi, Physics based deep learning for nonlinear two-phase flow
  in porous media, in: ECMOR XVII, Vol. 2020, European Association of
  Geoscientists \& Engineers, 2020, pp. 1--10.

\bibitem{WEN2021103223}
G.~Wen, M.~Tang, S.~M. Benson,
  \href{http://www.sciencedirect.com/science/article/pii/S1750583620306484}{Towards
  a predictor for co2 plume migration using deep neural networks},
  International Journal of Greenhouse Gas Control 105 (2021) 103223.
\newblock \href {https://doi.org/https://doi.org/10.1016/j.ijggc.2020.103223}
  {\path{doi:https://doi.org/10.1016/j.ijggc.2020.103223}}.
\newline\urlprefix\url{http://www.sciencedirect.com/science/article/pii/S1750583620306484}

\bibitem{wu2020physics}
H.~Wu, R.~Qiao, Physics-constrained deep learning for data assimilation of
  subsurface transport, Energy and AI 3 (2020) 100044.

\bibitem{he2020physics}
Q.~He, D.~Barajas-Solano, G.~Tartakovsky, A.~M. Tartakovsky, Physics-informed
  neural networks for multiphysics data assimilation with application to
  subsurface transport, Advances in Water Resources 141 (2020) 103610.

\bibitem{liu2020petrophysical}
M.~Liu, D.~Grana, Petrophysical characterization of deep saline aquifers for
  co2 storage using ensemble smoother and deep convolutional autoencoder,
  Advances in Water Resources 142 (2020) 103634.

\bibitem{jiang2021deep}
Z.~Jiang, P.~Tahmasebi, Z.~Mao, Deep residual u-net convolution neural networks
  with autoregressive strategy for fluid flow predictions in large-scale
  geosystems, Advances in Water Resources (2021) 103878.

\bibitem{haghighat2021sciann}
E.~Haghighat, R.~Juanes, Sciann: A keras/tensorflow wrapper for scientific
  computations and physics-informed deep learning using artificial neural
  networks, Computer Methods in Applied Mechanics and Engineering 373 (2021)
  113552.

\bibitem{Zhong2019}
Z.~Zhong, A.~Y. Sun, H.~Jeong, Predicting co2 plume migration in heterogeneous
  formations using conditional deep convolutional generative adversarial
  network, Water Resources Research 55~(7) (2019) 5830--5851.

\bibitem{haykin2010neural}
S.~Haykin, Neural Networks and Learning Machines, 3/E, Pearson Education India,
  2010.

\bibitem{Pruess1999}
K.~Pruess, C.~M. Oldenburg, G.~Moridis, Tough2 user's guide version 2, Tech.
  rep., Lawrence Berkeley National Lab.(LBNL), Berkeley, CA (United States)
  (1999).

\bibitem{eclipse}
{Schlumberger}, Eclipse reservoir simulation software reference manual (2014).

\bibitem{gccsidatabase}
{Global CCS Institute}, Ccs facilities database co2re, data retrieved from:
  \url{https://co2re.co/} (2020).

\bibitem{NationalAcademiesofSciencesEngineering2018}
NAS, {Negative Emissions Technologies and Reliable Sequestration}, 2018.
\newblock \href {https://doi.org/10.17226/25259} {\path{doi:10.17226/25259}}.

\bibitem{krevor2012relative}
S.~C. Krevor, R.~Pini, L.~Zuo, S.~M. Benson, Relative permeability and trapping
  of co2 and water in sandstone rocks at reservoir conditions, Water resources
  research 48~(2) (2012).

\bibitem{benson2013relative}
S.~Benson, R.~Pini, C.~Reynolds, S.~Krevor, Relative permeability analyses to
  describe multi-phase flow in co2 storage reservoirs, Global CCS Institute
  (2013).

\bibitem{tran2015learning}
D.~Tran, L.~Bourdev, R.~Fergus, L.~Torresani, M.~Paluri, Learning
  spatiotemporal features with 3d convolutional networks, in: Proceedings of
  the IEEE international conference on computer vision, 2015, pp. 4489--4497.

\bibitem{song2017end}
S.~Song, C.~Lan, J.~Xing, W.~Zeng, J.~Liu, An end-to-end spatio-temporal
  attention model for human action recognition from skeleton data, in:
  Proceedings of the AAAI conference on artificial intelligence, Vol.~31, 2017.

\bibitem{xie2018rethinking}
S.~Xie, C.~Sun, J.~Huang, Z.~Tu, K.~Murphy, Rethinking spatiotemporal feature
  learning: Speed-accuracy trade-offs in video classification, in: Proceedings
  of the European Conference on Computer Vision (ECCV), 2018, pp. 305--321.

\bibitem{he2016deep}
K.~He, X.~Zhang, S.~Ren, J.~Sun, Deep residual learning for image recognition,
  in: Proceedings of the IEEE conference on computer vision and pattern
  recognition, 2016, pp. 770--778.

\bibitem{ronneberger2015u}
O.~Ronneberger, P.~Fischer, T.~Brox, U-net: Convolutional networks for
  biomedical image segmentation, in: International Conference on Medical image
  computing and computer-assisted intervention, Springer, 2015, pp. 234--241.

\bibitem{kingma2014adam}
D.~P. Kingma, J.~Ba, Adam: A method for stochastic optimization, arXiv preprint
  arXiv:1412.6980 1 (2014).

\bibitem{glorot2010understanding}
X.~Glorot, Y.~Bengio, Understanding the difficulty of training deep feedforward
  neural networks, in: Proceedings of the thirteenth international conference
  on artificial intelligence and statistics, JMLR Workshop and Conference
  Proceedings, 2010, pp. 249--256.

\bibitem{van1995co2}
L.~Van~der Meer, The co2 storage efficiency of aquifers, Energy conversion and
  management 36~(6-9) (1995) 513--518.

\bibitem{osti_1177419}
R.~Okwen, S.~Frailey, H.~Leetaru, S.~Moulton, Assessing reservoir depositional
  environments to develop and quantify improvements in co2 storage efficiency.
  a reservoir simulation approach, Tech. rep., University of Illinois,
  Champaign, IL (United States) (2014).

\bibitem{gunter2004role}
W.~D. Gunter, S.~Bachu, S.~Benson, The role of hydrogeological and geochemical
  trapping in sedimentary basins for secure geological storage of carbon
  dioxide, Geological Society, London, Special Publications 233~(1) (2004)
  129--145.

\bibitem{gilfillan2009solubility}
S.~M. Gilfillan, B.~S. Lollar, G.~Holland, D.~Blagburn, S.~Stevens, M.~Schoell,
  M.~Cassidy, Z.~Ding, Z.~Zhou, G.~Lacrampe-Couloume, et~al., Solubility
  trapping in formation water as dominant co 2 sink in natural gas fields,
  Nature 458~(7238) (2009) 614--618.

\bibitem{suekane2008geological}
T.~Suekane, T.~Nobuso, S.~Hirai, M.~Kiyota, Geological storage of carbon
  dioxide by residual gas and solubility trapping, International Journal of
  Greenhouse Gas Control 2~(1) (2008) 58--64.

\bibitem{riaz2006onset}
A.~Riaz, M.~Hesse, H.~Tchelepi, F.~Orr, Onset of convection in a
  gravitationally unstable diffusive boundary layer in porous media, Journal of
  Fluid Mechanics 548~(1) (2006) 87--111.

\bibitem{macminn2012spreading}
C.~W. MacMinn, J.~A. Neufeld, M.~A. Hesse, H.~E. Huppert, Spreading and
  convective dissolution of carbon dioxide in vertically confined, horizontal
  aquifers, Water Resources Research 48~(11) (2012).

\bibitem{enick1990co2}
R.~M. Enick, S.~M. Klara, Co2 solubility in water and brine under reservoir
  conditions, Chemical Engineering Communications 90~(1) (1990) 23--33.

\bibitem{doughty2007modeling}
C.~Doughty, Modeling geologic storage of carbon dioxide: comparison of
  non-hysteretic and hysteretic characteristic curves, Energy Conversion and
  Management 48~(6) (2007) 1768--1781.

\bibitem{krevor2015capillary}
S.~Krevor, M.~J. Blunt, S.~M. Benson, C.~H. Pentland, C.~Reynolds,
  A.~Al-Menhali, B.~Niu, Capillary trapping for geologic carbon dioxide
  storage--from pore scale physics to field scale implications, International
  Journal of Greenhouse Gas Control 40 (2015) 221--237.

\bibitem{bachu1994aquifer}
S.~Bachu, W.~Gunter, E.~Perkins, Aquifer disposal of co2: hydrodynamic and
  mineral trapping, Energy conversion and management 35~(4) (1994) 269--279.

\bibitem{xu2004numerical}
T.~Xu, J.~A. Apps, K.~Pruess, Numerical simulation of co2 disposal by mineral
  trapping in deep aquifers, Applied geochemistry 19~(6) (2004) 917--936.

\bibitem{doecarbon}
U.~DOE, Carbon storage atlas--fifth edition (atlas v).[online]. national energy
  technical laboratory (2015).

\bibitem{USGS}
U.~G. S. G. C. D. S. R.~A. Team, \href{https://pubs.usgs.gov/ds/774/}{National
  assessment of geologic carbon dioxide storage resources—data (ver. 1.1,
  september 2013): U.s. geological survey data series 774, 13 p., plus 2
  appendixes and 2 large tables in separate files} (2013).
\newline\urlprefix\url{https://pubs.usgs.gov/ds/774/}

\bibitem{BOEM}
B.~of~Ocean Energy~Management,
  \href{https://www.data.boem.gov/Main/GandG.aspx}{Atlas of gulf of mexico gas
  and oil sands data}.
\newline\urlprefix\url{https://www.data.boem.gov/Main/GandG.aspx}

\bibitem{remy_boucher_wu_2009}
N.~Remy, A.~Boucher, J.~Wu, Applied Geostatistics with SGeMS: A User's Guide,
  Cambridge University Press, 2009.
\newblock \href {https://doi.org/10.1017/CBO9781139150019}
  {\path{doi:10.1017/CBO9781139150019}}.

\bibitem{carpentier2009conservation}
S.~Carpentier, K.~Roy-Chowdhury, Conservation of lateral stochastic structure
  of a medium in its simulated seismic response, Journal of Geophysical
  Research: Solid Earth 114~(B10) (2009).

\end{thebibliography}








\end{document}